\documentclass[11pt, a4paper]{article}

\usepackage{amssymb,amsmath,amsthm}
\usepackage{subfigure,color,colordvi,graphicx,xcolor}

\usepackage[only,llbracket,rrbracket,interleave]{stmaryrd}
\usepackage[colorlinks,bookmarksnumbered,bookmarksopen]{hyperref}

\usepackage{yhmath} 

\allowdisplaybreaks

\usepackage{xcolor}
\usepackage{ulem}

\newcommand{\bm}[1]{\text{\boldmath $#1$\unboldmath}}
\newcommand{\vect}[1]{\mathbf{#1}}
\newcommand{\mat}[1]{\mathbf{#1}}

\newcommand{\norm}[1]{\lVert#1\rVert}

\newcommand{\nsd}  {\ensuremath{\texttt{n}_{\texttt{sd}}}}
\newcommand{\numel}{\ensuremath{\texttt{n}_{\texttt{el}}}}
\newcommand{\nen}  {\ensuremath{\texttt{n}_{\texttt{en}}}}

\newcommand{\grad}{\bm{\nabla}}
\newcommand{\Div}{{\grad{\cdot}}}
\newcommand{\defo}{\grad^{\texttt{{s}}}}

\newcommand{\Insd}{\mat{I}}

\newcommand{\bu}{\bm{u}}
\newcommand{\bhu}{\bm{\hat{u}}}

\newcommand{\bhw}{\bm{\widehat{w}}}

\newcommand{\bx}{\bm{x}}
\newcommand{\bL}  {\bm{L}}
\newcommand{\bG}  {\bm{G}}
\newcommand{\bw}  {\bm{w}}

\newcommand{\bn}{\bm{n}}
\newcommand{\btau}{\bm{\tau}}

\newcommand{\bt}{\bm{t}}

\newcommand{\Ga}[1]{\Gamma_{\!\!#1}}

\newcommand{\nodalu}{\vect{u}}
\newcommand{\bxi}{\bm{\xi}}

\renewcommand{\SS}{\mathbb{S}}
\newcommand{\eltwo}{\ensuremath{\mathcal{L}_{2_{\!}}}}
\newcommand{\sobo}[1][1]{\ensuremath{\mathcal{H}^{#1\!}}}
\newcommand{\hDiv}[1]{\ensuremath{\mathcal{H}(\operatorname{div};{{#1}})}}

\newcommand{\Htrace}{\ensuremath{\mathcal{H}^{\frac{1}{2}}}}

\newcommand{\bigjump}[1]{\bigl\llbracket #1\bigr\rrbracket}

\newcommand{\bM}{\mat{M}}
\newcommand{\bD}{\vect{D}}
\newcommand{\bF}{\vect{F}}
\newcommand{\bU}{\vect{U}}

\usepackage{algorithm, algorithmic}

\usepackage{scalerel,stackengine}
\stackMath
\newcommand\reallywidehat[1]{%
	\savestack{\tmpbox}{\stretchto{%
			\scaleto{%
				\scalerel*[\widthof{\ensuremath{#1}}]{\kern-.6pt\bigwedge\kern-.6pt}%
				{\rule[-\textheight/2]{1ex}{\textheight}}
			}{\textheight}%
		}{0.5ex}}%
	\stackon[1pt]{#1}{\tmpbox}%
}

\graphicspath{{figsPDF/}{figsPNG/}}

\title{A conservative degree adaptive HDG method for transient incompressible flows}
\author{Agustina Felipe, Ruben Sevilla and Oubay Hassan}
\date{\small{Zienkiewicz Centre for Computational Engineering\\ Faculty of Science and Engineering, Swansea University, Wales, UK.}}

\begin{document}

\maketitle

\begin{abstract}
\noindent \textbf{Purpose:} This study aims to assess the accuracy of degree adaptive strategies in the context of incompressible Navier-Stokes flows using the high order hybridisable discontinuous Galerkin (HDG) method.

\noindent \textbf{Design/methodology/approach:} The work presents a series of numerical examples to show the inability of standard degree adaptive processes to accurate capture aerodynamic quantities of interest, in particular the drag. A new conservative projection is proposed and the results between a standard degree adaptive procedure and the adaptive process enhanced with this correction are compared. The examples involve two transient problems where flow vortices or a gust needs to be accurately propagated over long distances.

\noindent \textbf{Findings:} The lack of robustness and accuracy of a standard degree adaptive processes is linked to the violation of the free-divergence condition when projecting a solution from a space of polynomials of a given degree to a space of polynomials with a lower degree. Due to the coupling of velocity-pressure in incompressible flows, the violation of the incompressibility constraint leads to inaccurate pressure fields in the wake that have a sizeable effect on the drag. The new conservative projection proposed is found to remove all the numerical artefacts shown by the standard adaptive process.

\noindent \textbf{Originality/value:} This work proposes a new conservative projection for the degree adaptive process. The projection does not introduce a significant overhead because it requires to solve an element-by-element problem and only for those elements where the adaptive process lowers the degree of approximation. Numerical results show that with the proposed projection non-physical oscillations in the drag disappear and the results are in good agreement with reference solutions. 


\end{abstract}

\section{Introduction}  \label{sc:intro}

The accurate simulation of transient incompressible fluid flows is a central challenge in many computational fluid dynamics (CFD) applications, including areas such as civil, aerospace, chemical and biomedical engineering. From a numerical point of view, several difficulties arise when solving the incompressible Navier-Stokes equations due to their non-linear nature and the intricate coupling between velocity and pressure fields~\cite{quartapelle2013numerical}. When unsteady phenomena are of interest an extra difficulty is to accurate propagate vortices over long distances. 

High-order methods are attractive for the simulation of transient flows due to the lower dissipation and dispersion errors, when compared to low order methods~\cite{ekaterinaris2005high,ainsworth2006dispersive,wang2013high}. Continuous and discontinuous Galerkin (DG) methods have their own advantages and disadvantages and have both been successfully applied to a variety of problems in CFD~\cite{chalot2010higher,chalot2015higher,gross2015automatic,sevilla2013analysis,bassi2007implicit,liu2000high,montlaur2010discontinuous,ferrer2011high,lehrenfeld2016high}. Two properties that make DG a preferred option in some cases is the ability to easily handle a variable degree of approximation and the easier definition of the required stabilisation for convection dominated flows~\cite{kompenhans2016comparisons,ekelschot2017p,paipuri2018comparison}. The main disadvantage of DG methods is commonly attributed to the duplication of degrees of freedom~\cite{kirby2012cg,yakovlev2015cg}, which in turns is the property that facilitates the implementation of variable degree of approximation. 

The hybridisable discontinuous Galerkin (HDG) method, originally proposed by Cockburn and co-workers~\cite{Jay-CG:05,cockburn2009derivation} employs hybridisation to reduce the number of coupled degrees of freedom and has become popular for CFD applications. With HDG, it is possible to use approximations of equal order for both velocity and pressure, circumventing the Ladyzhenskaya-Babu\v{s}ka-Brezzi (LBB) condition. From a computational perspective, the size of the global problem only involves the mean value of the pressure in each element even for high-order approximations, reducing even further the size of the global system of equations to be solved. Furthermore, an important advantage of HDG is the ability to build a super-convergent velocity field~\cite{cockburn2013conditions}. The development and application of HDG methods to incompressible flows include the solution of Stokes flows~\cite{Nguyen-CNP:10,Nguyen-NPC:10,cockburn2014devising,cockburn2013conditions,giacomini2018superconvergent} and the incompressible Navier-Stokes equations~\cite{Nguyen-NPC:11,giorgiani2014hybridizable,rhebergen2018hybridizable,gurkan2019extended}. 

To accurately and efficiently capture transient flow phenomena, mesh adaptation techniques are traditionally employed in a low order context. For high-order methods the use of degree adaptivity offers a new alternative to provide the required accuracy only in the regions of the domain where is needed, minimising the computational overhead of high-order approximations and circumventing the need to modify the mesh topology. In the context of HDG, the use of mesh and degree adaptivity has been considered for a variety of problems, including incompressible flows~\cite{giorgiani2014hybridizable,leng2021adaptive}. In HDG methods, the ability to build a super-convergent solution can be used to devise a cheap error indicator to drive the adaptivity. This strategy was first exploited in~\cite{giorgiani2013hybridizable} for wave propagation problems.

This work considers the solution of the incompressible Navier-Stokes equations using a degree adaptive HDG method. First, it is shown that a degree adaptive process can lead to unphysical oscillations in aerodynamic quantities of interest, especially the drag, if the adaptive process reduces the degree of approximation during the time marching process. This phenomenon is linked to the violation of the free-divergence condition during the projection of the solution from a space of polynomials of degree $r$ to a space of polynomials of degree $s$, with $s<r$. Second, this work proposes a conservative projection to guarantee mass conservation during the projection stage. The proposed projection does not introduce a significant overhead because it induces the solution of an element-by-element problem and only for those elements where the adaptive process lowers the degree of approximation. Numerical examples are used to illustrate the benefits of the proposed conservative projection using two dimensional examples.

The remainder of the paper is organised as follows. Section~\ref{sc:HDG_INS} briefly summarises the numerical solution of the incompressible Navier-Stokes using the HDG method. In Section~\ref{sc:degreeAdaptive} the degree adaptive strategy proposed in this work is outlined, including the proposed conservative projection to guarantee mass conservation. Section~\ref{sc:examples} present numerical examples to illustrate the effect of using a standard adaptive process that violates the free-divergence condition during the projection stage and the benefits of the proposed conservative projection. Finally, the conclusions of the work are presented in Section~\ref{sc:conclusions}.

\section{HDG solution of the incompressible Navier-Stokes equations} \label{sc:HDG_INS}

This section summarises the HDG formulation employed to numerically solve the transient incompressible Navier-Stokes equations. Except from the addition of the transient term, the formulation follows the work in~\cite{Tutorial-GSH:2020}, so only the main ingredients required to present the proposed degree adaptive strategy are considered here.

\subsection{HDG formulation} \label{sc:HDGformulation}

The strong form of the transient incompressible Navier-Stokes equations in an open bounded domain $\Omega \subset \mathbb{R}^{\nsd}$, where $\nsd$ is the number of spatial dimensions, is written as
\begin{equation} \label{eq:NS}
\left\{\begin{aligned}
	\bu_t -\grad\cdot(2\nu\defo{\bu} - p \Insd) + \grad {\cdot} (\bu \otimes \bu) &= \bm{s}       &&\text{in $\Omega \times (0,T]$,}\\
	\Div\bu &= 0  &&\text{in $\Omega \times (0,T]$,}\\
	\bu &= \bu_D  &&\text{on $\Ga{D} \times (0,T]$,}\\
	\bigl((2\nu\defo{\bu} - p \Insd) - (\bu \otimes \bu)\bigr)\,\bn &= \bt        &&\text{on $\Ga{N} \times (0,T]$.}\\
	\bu &= \bu_0  &&\text{in $\Omega \times \{0\}$,}\\
\end{aligned}\right.
\end{equation}
where the $\bu$ is the velocity vector, $p$ is the pressure, $\nu$ is the kinematic viscosity, $\defo{\bu} := ( \grad\bu + \grad^T\bu)/2$ is the strain-rate tensor, $\bm{s}$ is the source term, $\bu_D$ is the imposed velocity on the Dirichlet part of the boundary, $\Ga{D}$, $\bt$ is the imposed traction on the Neumann part of the boundary, $\Ga{N}$, $\bn$ is the outward unit normal vector to the boundary, $\bu_0$ is the initial condition and $T$ denotes the final time.

The HDG method uses a mixed formulation leading to a rewriting of the momentum equation as a first-order partial differential equation, namely
\begin{equation} \label{eq:NS}
\bu_t + \Div(\sqrt{2\nu}\bL + p \Insd) + \grad {\cdot} (\bu \otimes \bu) = \bm{s}   \quad    \text{in $\Omega \times (0,T]$,}
\end{equation}
where $\bL= - \sqrt{2\nu} \defo\bu$ is the so-called mixed variable.

After discretising the domain in a set of $\numel$ non-overlapping elements $\Omega_e$, the mixed problem is written in each element and interface conditions to enforce the continuity of the solution and the continuity of the fluxes are introduced~\cite{Tutorial-GSH:2020}. A distinctive feature of the HDG method is the introduction of the trace of the velocity, also called hybrid velocity, as an independent variable on the mesh skeleton, defined as
\begin{equation}\label{eq:Gamma}
\Gamma := \Big[ \bigcup_{e=1}^{\numel} \partial\Omega_e \Big]\setminus\partial\Omega.
\end{equation}

The resulting problem is then solved in two stages. First the element-by-element local problems define a pure Dirichlet problem
\begin{equation} \label{eq:strongLocal}
\left\{\begin{aligned}
	\bL + \sqrt{2\nu} \defo{\bu}                  &= \bm{0}   &&\text{in $\Omega_e$,} \\	
	\bu_t + \Div(\sqrt{2\nu} \bL + p \Insd) + \grad {\cdot} (\bu \otimes \bu) &= \bm{s}   &&\text{in $\Omega_e$,} \\
	\Div \bu                                          &= 0           &&\text{in $\Omega_e$,} \\
	\bu                                                           &= \bu_D   &&\text{on $\partial\Omega_e \cap \Ga{D}$,}\\
	\bu                                                           &= \bhu     &&\text{on $\partial\Omega_e \setminus \Ga{D}$,}\\
	\bu &= \bu_0  &&\text{in $\Omega_e \times \{0\}$,}\\
	\langle p, 1 \rangle_{\partial\Omega_e} & = \rho_e, \\
\end{aligned} \right.
\end{equation}
where the last equation is required to remove the indeterminacy of the pressure, $\langle \cdot, \cdot \rangle_{S}$ denotes the classical $\eltwo$ inner product for vector-valued functions on $S \subset\Gamma\cup\partial\Omega$ and $\bhu$ is the hybrid velocity.

Second, the global problem is given by
\begin{equation} \label{eq:strongGlobal}
\left\{\begin{aligned}
	\bigjump{\bigl((\sqrt{2\nu} \bL + p \Insd)+ (\bu \otimes \bu)\bigr)\,\bn} &= \bm{0}  &&\text{on $\Gamma$,}\\
	\bigl((\sqrt{2\nu} \bL + p \Insd)+ (\bu \otimes \bu)\bigr)\,\bn  &= -\bt       &&\text{on $\Ga{N}$,}\\
	\langle \bhu \cdot \bn_e , 1 \rangle_{\partial \Omega_e \setminus \Ga{D}}  
	+ \langle \bu_D \cdot \bn_e, 1 \rangle_{\partial \Omega_e\cap \Ga{D}} & = 0 ,
\end{aligned} \right.
\end{equation}
where the last equation is induced by the free-divergence condition in the local problems. It is worth noting that there is no need to impose the continuity of the solution in the global problem due to the unique definition of the hybrid velocity in each face of the mesh skeleton and the imposition of $\bu = \bhu$ in the local problems~\eqref{eq:strongLocal}.

\subsection{Weak forms and the HDG stabilisation} \label{sc:HDGweak}

For each element, the weak formulation of local problems can be written as is as follows: find $(\bL_e, \bu, p) \in \left[\hDiv{\Omega};\SS\right] \times \left[\sobo(\Omega)\right]^{\nsd} \times \sobo(\Omega)$ such that
\begin{equation} \label{eq:local}
\left\{
	\begin{aligned} 
		{-}&\bigl( \bG, \bL \bigr)_{\Omega_e}  
		+ \bigl( \Div(\sqrt{2\nu}\bG), \bu \bigr)_{\Omega_e}
		\\[-0.5ex] &\hspace{65pt}
		= \langle \bG\,\bn, \sqrt{2\nu}\,\bu_D \rangle_{\partial\Omega_e\cap\Ga{D}} 
		+ \langle \bG\,\bn, \sqrt{2\nu}\,\bhu \rangle_{\partial \Omega_e \setminus \Ga{D} } ,
		\\[1ex]
		&\bigl( \bw, \bu_t \bigr)_{\Omega_e} {+} \bigl( \bw, \grad{\cdot}( \sqrt{2\nu} \bL) \bigr)_{\Omega_e}  
		{+} \bigl( \bw, \grad p \bigr)_{\Omega_e}  
		\\[-0.5ex] &\hspace{65pt}
		{+} \bigl\langle \bw, (\reallywidehat{\sqrt{2\nu} \bL {+} p \Insd})\,\bn{-}(\sqrt{2\nu} \bL {+} p \Insd)\,\bn \bigr\rangle_{\partial\Omega_e}
		\\[-0.5ex] &\hspace{75pt}
		{-} \bigl( \grad\bw , \bu \otimes \bu \bigr)_{\Omega_e} 
		{+} \langle \bw,(\widehat{\bu \otimes \bu})  \bn \rangle_{\partial\Omega_e} 
		= \bigl( \bw, \bm{s} \bigr)_{\Omega_e} ,
		\\[1ex]
		&\bigl( \grad q, \bu \bigr)_{\Omega_e}  
		= \langle q, \bu_D \cdot \bn \rangle_{\partial \Omega_e \cap \Ga{D}}
		+ \langle q, \bhu \cdot \bn \rangle_{\partial \Omega_e \setminus \Ga{D} } ,
		\\
		&\langle p, 1 \rangle_{\partial\Omega_e}   = \rho_e ,
	\end{aligned}\right.
\end{equation}
for all $(\bG, \bw, q) \in \left[\hDiv{\Omega_e};\SS\right] \times \left[\sobo(\Omega_e)\right]^{\nsd} \times \sobo(\Omega_e)$, where $\left[\hDiv{\Omega_e};\SS\right]$ denotes the space of square-integrable symmetric tensors $\SS$ of order $\nsd$ with square-integrable row-wise divergence. 

Similarly, the weak form of the global problem reads: find $\bhu \in \left[\Htrace(\Gamma \cup \Ga{N})\right]^{\nsd}$ and $\bm{\rho} \in \mathbb{R}^{\numel}$ that satisfies 
\begin{equation}\label{eq:global}
	\left\{\begin{aligned}
		\sum_{e=1}^{\numel} \Big\{ 
		\bigl\langle\bhw , (\reallywidehat{\sqrt{2\nu}\bL{+}p\Insd})\,\bn
		{+} (\widehat{\bu {\otimes} \bu})\,\bn_e \bigr\rangle_{\partial\Omega_e\setminus\partial\Omega} 
		\hspace{100pt} & \\[-1.5ex] 
		{+} \bigl\langle \bhw , (\reallywidehat{\sqrt{2\nu}\bL{+}p\Insd})\,\bn
		{+} (\widehat{\bu {\otimes} \bu})\,\bn_e {+} \bt \bigr\rangle_{\partial \Omega_e \cap \Ga{N}}  \Big\}
		{=} 0 , &
		\\
		\langle \bhu \cdot \bn_e , 1 \rangle_{\partial \Omega_e \setminus \Ga{D} } 
		= - \langle \bu_D  \cdot \bn_e , 1 \rangle_{\partial \Omega_e \cap \Ga{D}} \qquad \text{for $e=1,\dotsc,\numel$}, &
	\end{aligned}\right.
\end{equation}
for all $\bhw \in \left[\eltwo(\Gamma \cup \Ga{N})\right]^{\nsd}$. 

Following~\cite{Jay-CGL:09,RS-SH:16,Tutorial-GSH:2020}, the numerical traces appearing in the local and global problems are defined as
\begin{subequations}\label{eq:NumFlux}
	\begin{equation} \label{eq:traceDiffusion}
		\hspace{-1em}
		(\reallywidehat{\sqrt{2\nu} \bL {+} p \Insd})\,\bn {:=} 
		\hspace{-0.25em}
		\begin{cases}
			(\sqrt{2\nu} \bL {+} p \Insd)\,\bn {+} \tau^d (\bu {-} \bu_D) &\hspace{-0.75em} \text{on $\partial\Omega_e\cap\Ga{D}$,} \\
			(\sqrt{2\nu} \bL {+} p \Insd)\,\bn {+} \tau^d (\bu {-} \bhu) &\hspace{-0.75em} \text{elsewhere,}  
		\end{cases}
	\end{equation}
	\begin{equation} \label{eq:traceAdvection}
		(\widehat{\bu \otimes \bu})\bn_e := \begin{cases}
			(\bhu \otimes \bu_D)\bn + \tau^a(\bu-\bu_D) & \text{on $\partial\Omega_e\cap\Ga{D}$,} \\
			(\bhu \otimes \bhu)\bn + \tau^a(\bu-\bhu) & \text{elsewhere,}  
		\end{cases}
	\end{equation}
\end{subequations}
where $\tau^d$ and $\tau^a$ are the diffusive and convective stabilisation parameters, respectively, which in this work are defined as
\begin{equation}\label{eq:tauDif}
	\tau^d = 10 {\nu}/{\ell} , \qquad \tau^a = \max_{\bx} \norm{\bu(\bx)}_2,
\end{equation}
where the factor 10 in the diffusive stabilisation is taken following previous work on HDG methods~\cite{giacomini2018superconvergent,Tutorial-GSH:2020} and the maximum in the convective stabilisation is taken over all the mesh nodes. Other options for the convective stabilisation, not considered here, have been recently proposed in~\cite{vieira2024face}.

The selected parameters ensure the satisfaction of the admissibility condition
introduced in~\cite{Jay-CGL:09} to guarantee stability and well-posedness,
\begin{equation}
	\min _{\bx \in \partial \Omega_{e}}\left\{\tau^{d}+\tau^{a}-\hat{\bu} \cdot \bm{n}\right\} \geq \gamma>0
\end{equation}
for some constant $\gamma$.

\subsection{HDG solution algorithm} \label{sc:HDGsolution}

Introducing the numerical traces~\eqref{eq:NumFlux} into the local problems leads to the following residuals
\begin{equation}\label{eq:localResiduals}
\begin{aligned} 
\mathcal{R}^e_1  :&=  \bigl( \bG, \bL_e \bigr)_{\Omega_e}  
{-} \bigl( \Div(\sqrt{2\nu}\bG), \bu_e \bigr)_{\Omega_e}
{-} \langle \bG\,\bn_e, \sqrt{2\nu}\,\bu_D \rangle_{\partial\Omega_e\cap\Ga{D}} 
\\ & {+} \langle \bG\,\bn_e, \sqrt{2\nu}\,\bhu \rangle_{\partial \Omega_e \setminus\Ga{D} },
\\	
\mathcal{R}^e_2  :&= \bigl( \bw, \bu_t \bigr)_{\Omega_e}  {+} \bigl( \bw, \grad{\cdot}( \sqrt{2\nu} \bL) \bigr)_{\Omega_e}  
{+} \bigl( \bw, \grad p \bigr)_{\Omega_e} 
{-} \bigl( \grad\bw , \bu \otimes \bu \bigr)_{\Omega_e} \\
& {+} \langle \bw , \tau \bu \rangle_{\partial\Omega_e} 
- \bigl( \bw, \bm{s} \bigr)_{\Omega_e}
{-} \langle \bw, (\tau {-} \bhu{\cdot}\bn) \bu_D \rangle_{\partial\Omega_e \cap \Ga{D}}
\\ &{-} \langle \bw, (\tau {-} \bhu{\cdot}\bn) \bhu \rangle_{\partial\Omega_e \setminus \Ga{D}},
\\
\mathcal{R}^e_3  :&=\bigl( \grad q, \bu_e \bigr)_{\Omega_e}  
{-} \langle q, \bu_D \cdot \bn_e \rangle_{\partial \Omega_e \cap \Ga{D}}
{-} \langle q, \bhu \cdot \bn_e \rangle_{\partial \Omega_e \setminus \Ga{D} } ,
\\
\mathcal{R}^e_4  :&= \langle p_e, 1 \rangle_{\partial\Omega_e}   - \rho_e,
\end{aligned}
\end{equation}
where $\tau = \tau^d + \tau^a$. Similarly, the global problem leads to the residuals
\begin{equation}\label{eq:globalResiduals}
\begin{aligned}
\mathcal{R}_5 :&= \sum_{e=1}^{\numel} \Big\{ 
\langle\bhw , (\sqrt{2\nu}\bL_e{+}p_e\Insd)\,\bn_e \rangle_{\partial \Omega_e\setminus\Ga{D}}    
{+} \langle\bhw , \tau\bu_e \rangle_{\partial \Omega_e\setminus\Ga{D}}    
\\
&{-} \langle \bhw , \tau \bhu \rangle_{\partial \Omega_e\cap\Gamma} {-} \langle \bhw , (\tau {-} \bhu {\cdot} \bn_e)\bhu \rangle_{\partial \Omega_e\cap\Ga{N}}  
{+} \langle \bhw , \bt \rangle_{\partial \Omega_e \cap \Ga{N}}  \Big\},
\\
\mathcal{R}^e_6 :&= \langle \bhu \cdot \bn_e , 1  \rangle_{\partial \Omega_e \setminus \Ga{D} }  
=   -\langle \bu_D  \cdot \bn_e , 1 \rangle_{\partial \Omega_e \cap \Ga{D}}.
\end{aligned}
\end{equation}

In this work, the spatial discretisation is performed using isoparametric elements, including curved elements in the vicinity of curved boundaries. The approximation for the velocity $\bu$, pressure $p$, mixed variable $\bL$ is defined in a reference element, with polynomials of order $k \geq 1$. Similarly, the approximation of the hybrid velocity $\bhu$ is defined on a reference face, with polynomials of order $\hat{k} \geq 1$. The focus of this work is on degree adaptivity and, therefore, the current implementation supports an arbitrary order of approximation on each element. When two neighbouring elements use two different orders, the approximation of the hybrid velocity on the shared face between the two elements takes the maximum of the orders used on each element. This choice for the order of approximation of the hybrid velocity guarantees the optimal convergence properties of the HDG method with variable degree of approximation~\cite{chen2012analysis,chen2014analysis}.

The temporal discretisation is performed using high-order explicit first stage singly diagonal implicit Runge-Kutta (ESDIRK) integration methods. More precisely the fourth-order six-stage (ESDIRK46) scheme proposed in~\cite{kennedy2016diagonally} is utilised in all the numerical examples. ESDIRK methods retain the stability properties of implicit Runge-Kutta methods and provide improved performance when compared to singly-diagonal implicit Runge-Kutta methods. In addition ESDIRK  methods have been found to be more computationally efficient than other single-stage low order implicit schemes such as backward differentiation formulae (BDF) methods~\cite{bijl2001time}.

To strongly enforce the symmetry of the stress tensor, the present work considers the so-called Voigt notation, which has been shown~\cite{HDG-Elasticity2018,giacomini2018superconvergent,Tutorial-GSH:2020} to provide the super-convergent properties described in the next section and extra efficiency when compared to formulations where the symmetry is not strongly enforced. 

As usual in an HDG context~\cite{cockburn2009derivation,Nguyen-NPC:09-LinearConDif,Nguyen-NPC:09-NonLinearConDif,Nguyen-NPC:11-NS}, hybridisation is used and in each Newton-Raphson iteration, a global problem is solved to obtain the hybrid velocity and the mean pressure, followed by the solution of multiple local problems, element-by-element, to obtain the velocity, pressure and mixed variable in each element.

For a more detailed presentation of the HDG formulation and its implementation, the reader is referred to~\cite{RS-SH:16,Tutorial-GSH:2020,HDGlab-GSH-20,JVP_HDG-VGSH:20}. For a more detailed description of the Newton-Raphson linearisation strategy the reader is referred to~\cite{Sanjay2020,vieira2024face}.

\section{Degree adaptive strategy} \label{sc:degreeAdaptive}

This works exploits the ability of the HDG method to build a cheap error indicator using the a super-convergent approximation of the velocity field. In this section the strategy to build the super-convergent velocity and the error indicator are briefly recalled, before presenting the proposed correction to guarantee conservation in a transient degree adaptive process.

\subsection{Super-convergent postprocess of the velocity} \label{sc:postprocess}

An attractive property of HDG methods is the possibility to construct a super-convergent approximation of the velocity field, also called the postprocessed velocity and denoted by $\bu^\star$, by solving the element-by-element problem defined as
\begin{equation}\label{eq:postprocess}
\left\{\begin{aligned}
	\grad {\cdot} \left(\sqrt{2\nu} \defo{\bu^\star} \right) & = - \grad {\cdot} \bL , &&\text{in $\Omega_e$,}\\
	\left( \sqrt{2\nu} \defo{\bu^\star} \right) \bn & = - \bL \bn, &&\text{on $\partial \Omega_e$,}\\
	( \bu^\star, 1 )_{\Omega_e} & = ( \bu , 1)_{\Omega_e} , &&\\
	( \grad \times \bu^\star , 1 )_{\Omega_e}  & =  \langle \bu_D \cdot \btau , 1 \rangle_{\partial \Omega_e \cap \Gamma_D} + \langle \bhu \cdot \btau , 1 \rangle_{\partial \Omega_e \setminus \Gamma_D}, &&\\
\end{aligned}\right.
\end{equation}
where $\btau$ is the tangential direction to the boundary.

The first equation in~\eqref{eq:postprocess} is obtained after applying the divergence operator to the equation that defines the mixed variable and the boundary condition imposes equilibrated fluxes on the boundary of each element. The two last equations in~\eqref{eq:postprocess} are introduced to remove the indeterminacy associated with the translational and rotational modes.

Previous work on HDG methods~\cite{cesmelioglu2017analysis} have proved that if the velocity, pressure and mixed variable are approximated with polynomials of degree $k \geq 1$, their respective errors, measured in the $\eltwo(\Omega)$ norm, converge with order $k+1$ whereas the postprocessed velocity has an error that converges with order $k+2$, at least in diffusion dominated areas.

\subsection{Error indicator} \label{sc:error}

The possibility to build a super-convergent velocity in HDG method was first exploited in~\cite{giorgiani2013hybridizable} to devise a cheap error indicator to drive a degree adaptive process in wave propagation problems. This strategy has also been used for incompressible Navier-Stokes flows~\cite{giorgiani2014hybridizable}, Stokes flows~\cite{sevilla2018hdg} and linear elastic problems in~\cite{sevilla2019hdg}.

The main idea consists of approximating the error in the velocity field, $\bu$, in an element, $\Omega_e$, as
\begin{equation}\label{eq:errorVelo}
E_e = \left[  \frac{1}{\left|\Omega_e\right|} \int_{\Omega_e} \left( \bu -\bu^\star \right) \cdot \left( \bu -\bu^\star \right) d\Omega \right]^{1/2},
\end{equation}
where the normalisation using the element measure is crucial for meshes with large variation in element size~\cite{diez1999unified}.

The procedure to adapt the degree of approximation aims at ensuring that the error in each element is below a user-defined tolerance $\varepsilon$~\cite{remacle2003adaptive}. The degree is iteratively adapted as $k_e^r = k_e^{r-1} + \Delta k_e$ where $r$ denotes the degree adaptive iteration and the increment is given by
\begin{equation}\label{eq:deltaK}
\Delta k_e = \left\lceil \log_{10} \left(\frac{\varepsilon}{E_e} \right)\right\rceil,
\end{equation}
where $\lceil \cdot \rceil$ denotes the ceiling function. The base 10 in the logarithm base has been selected to minimise the number of iterations required in the degree adaptive process, but higher values can be used for a less aggressive adaptation~\cite{fidkowski2007triangular,giorgiani2014hybridizable}. The user defined tolerance $\varepsilon$ could be selected to be a piecewise constant function with different values in different elements/regions, but for simplicity, a constant value is used in this work and detailed for each example.

\subsection{Conservative projection for transient problems} \label{sc:conservativeProj}

For steady problems the adaptive process starts computing the solution for a given degree of approximation, commonly $k=1$ in all elements. After the solution is computed, the postprocessed velocity and the error indicator are evaluated element-by-element using \eqref{eq:postprocess} and \eqref{eq:errorVelo}, respectively. With this information, a new elemental degree map is defined~\eqref{eq:deltaK}. The process is repeated with the new elemental degree map until the error provided by the error indicator in each element is below the user-defined tolerance. 

A solution computed with a given degree map can be used to build a better initial guess of the Newton-Raphson scheme by interpolating the solution at the new nodal distribution within each element. Let us consider that the solution in one element has been computed using a polynomial approximation of degree $r$ and the new degree to be used in the element is $s$. The solution is initially approximated as
\begin{equation}\label{eq:degreeR}
\bu^r(\bxi) =  \sum_{j=1}^{\nen^r} \nodalu_j^r N_j^r(\bxi),
\end{equation}
where $\nen^r$ denotes the number of element nodes, $\nodalu_j$ are the nodal values of the solution and $N_j^r$ are the polynomial shape functions of degree $r$ defined, on a reference element, from the set of nodes $\{\bxi^r\}_{i=1,\ldots,\nen^r}$. The interpolation in the new set of nodes associated to a degree $s$, $\{\bxi^s\}$, can be written as
\begin{equation}\label{eq:degreeSProj}
\bu^s(\bxi) =  \sum_{j=1}^{\nen^s} \nodalu_j^s N_j^s(\bxi),
\end{equation}
where $\nodalu_j^s = \bu^r(\bxi^s_j)$.

A crucial difference of a degree adaptive process for transient problems, compared to the steady case, is that the projection of the solution at time $t^n$ to the desired degree map is required to compute the solution at time $t^{n+1}$ and the projection is not just used as an initial guess of the Newton-Raphson scheme. Let us consider the case where the solution in one element at time $t^n$ is computed with a degree $r$ and the degree adaptive process changes the required degree in the element to be $s$. The projection given by~\eqref{eq:degreeSProj} does not generally guarantee that the projected velocity field at time $t^n$ is divergence-free. More precisely, if $s \geq r$, i.e. if the adaptive process increases or maintains the degree of approximation  in the element, the projection does not change the velocity field at time $t^n$ because the space of polynomials of degree $r$ is a subset of the space of polynomials of degree $s$. However, if $s<r$, i.e. if the adaptive process decreases the degree of approximation in the element, the projection changes the velocity field at time $t^n$ and the incompressibility constraint is, in general, violated.

To avoid this problem, this work proposes a new projection based on the constrained minimisation problem
\begin{equation} \label{eq:conservativeProj}
\left\{	\begin{aligned}
	\min_{\bu^s_j} \quad & \int_{\Omega_e} \left( \bu^s -\bu^r \right) \cdot \left( \bu^s -\bu^r \right) d\Omega \\
	\textrm{s.t.} \quad & \int_{\partial\Omega_e} \bu^s \cdot \bn \, d\Gamma = 0 \\
\end{aligned}
\right.
\end{equation}

The discrete version of the minimisation problem is a classical $\mathcal{L}^2(\Omega_e)$ projection of the solution, whereas the constraint is imposed using a Lagrange multiplier. The resulting system of linear equations to be solved in an element where the adaptive process decreases the degree of approximation can be written as
\begin{equation} \label{eq:conservativeProjDiscrete}
\begin{bmatrix}
	\bM & \mat{0} & \bD_1\\
	\mat{0} & \bM & \bD_2\\
	\bD_1^T & \bD_2^T & 0
\end{bmatrix}
\begin{bmatrix}
	\bU_1^s \\
	\bU_2^s  \\
	\vect{\lambda}
\end{bmatrix}
=
\begin{bmatrix}
	\bF_1 \\
	\bF_2 \\
	0
\end{bmatrix},
\end{equation}
in two dimensions, where $\bU_a^s$ is the vector containing the nodal values of the $a$-th component of the projected free-divergence velocity field, $\lambda$ is the Lagrange multiplier,
\begin{equation} \label{eq:conservativeProjMats}
M_{ij} := \int_{\Omega_e} N_i N_j\, d\Omega,  
\quad
(D_a)_i := \int_{\partial \Omega_e} N_i n_a\, d\Gamma,
\quad
(F_a)_i := \int_{\Omega_e} N_i u_a^r\, d\Omega,
\end{equation}
and $u_a^r$ is the $a$-th component of the original velocity field, approximated with polynomials of degree $r$.

It is worth emphasising that the minimisation problem, i.e. the solution of the linear system~\eqref{eq:conservativeProjMats}, is only required on those elements where the adaptive process decreases the degree of approximation and the size of the linear system in two dimensions is $2\nen+1$ where $\nen$ is the number of element nodes. In addition, the problem is solved independently on each element so it can be trivially parallelised to minimise the computational overhead.

Algorithm~\ref{alg:HDGadaptivity} describes the degree adaptive process, including the proposed conservative projection, where $\texttt{n}_\texttt{steps}$ is the number of time steps, $\texttt{n}_\texttt{adaptivity}$ is the number of times the adaptivity is repeated each time step and $\texttt{n}_\texttt{NR}$ is the maximum number of iterations used in the Newton-Raphson scheme.
\begin{algorithm}[!tb]
	\caption{Degree adaptive HDG method}
	\label{alg:HDGadaptivity}
	\begin{algorithmic}[1]
		\STATE Initialise polynomial degree map $\{k_e\}_{e = 1, \ldots, \numel}$
		\STATE Set desired error $\varepsilon$
		\FOR{$i_\texttt{s} \gets 1$ to $\texttt{n}_\texttt{steps}$}  
		\FOR{$i_\texttt{a} \gets 1$ to $\texttt{n}_\texttt{adaptivity}$} 
		\FOR{$i_\texttt{NR} \gets 1$ to $\texttt{n}_\texttt{NR}$}  
		\STATE Solve global problem, linearising the residuals of Equation~\eqref{eq:globalResiduals}
		\STATE Solve local problem, linearising the residuals of Equation~\eqref{eq:localResiduals}
		\ENDFOR
		\FOR{$i_\texttt{el} \gets 1$ to $\numel$} 
		\STATE Compute super-convergent velocity using Equation~\eqref{eq:postprocess}
		\STATE Compute error indicator using Equation~\eqref{eq:errorVelo}
		\STATE Update the degree using Equation~\eqref{eq:deltaK}
		\IF{$\Delta k_e < 0$}
		\STATE Compute conservative projection using Equation~\eqref{eq:conservativeProjDiscrete}
		\ENDIF
		\ENDFOR
		\ENDFOR
		\ENDFOR		
	\end{algorithmic}
\end{algorithm}
In the current implementation the maximum number of iterations of the Newton-Raphson scheme is five, but given the quadratic convergence only an average of three iterations are needed to reach the desired tolerance, set to $10^{-10}$ for all the residuals of the global and local problems. Given the large time steps used in the time marching process, numerical examples will be used to show that two adaptive iterations per time steps are required to obtain a converged solution with the desired error in each time step.

\section{Numerical Results} \label{sc:examples}

This section presents four numerical examples. The first two examples are used to verify the optimal convergence properties of the method in terms of both the spatial and temporal discretisation. The last two examples illustrate the benefits of the proposed conservative projection within a degree adaptive process. The proposed approach is compared to a standard degree adaptive process and to an adaptive process where the degree of approximation is not allowed to be lowered during the time marching. In both examples, reference solutions using a uniform degree of approximation are used to quantify the extra accuracy provided by the proposed conservative projection.

\subsection{Kovasznay flow} \label{sc:Kovasznay}

The first example considers the Kovasznay flow~\cite{Kovasznay1948}, which provides an analytical solution of the incompressible Navier-Stokes equations. The computational domain is a unit square, $\Omega=[0,1]^2$, and the analytical solution is given by
\begin{equation}
	\bu(\bx) = \begin{Bmatrix}
		1-\exp(2\lambda x_1)\cos(2\pi x_2) \\
		\frac{\lambda}{2\pi}\exp(2\lambda x_1)\sin(2\pi x_2)
	\end{Bmatrix}, 
	\quad
	p(\bx) = -\frac{1}{2}\exp(4\lambda x_1) + C,
	\label{eq:Kovasznay}
\end{equation}
where $\lambda = \dfrac{{Re}}{2} - \sqrt{\dfrac{{Re}^2}{4} + 4\pi^2}$ and $C = \dfrac{1}{8}\left[1 + \exp(4\lambda) - \dfrac{1}{2\lambda}(1 - \exp(4\lambda))\right]$. 

A Neumann boundary condition, corresponding to the exact solution, is imposed on the bottom part of the boundary, whereas Dirichlet boundary conditions, corresponding to the exact velocity, are imposed on the rest of the boundary.

\begin{figure}[!tb]
	\centering
	\subfigure[Mesh 1]{\includegraphics[width=0.3\textwidth]{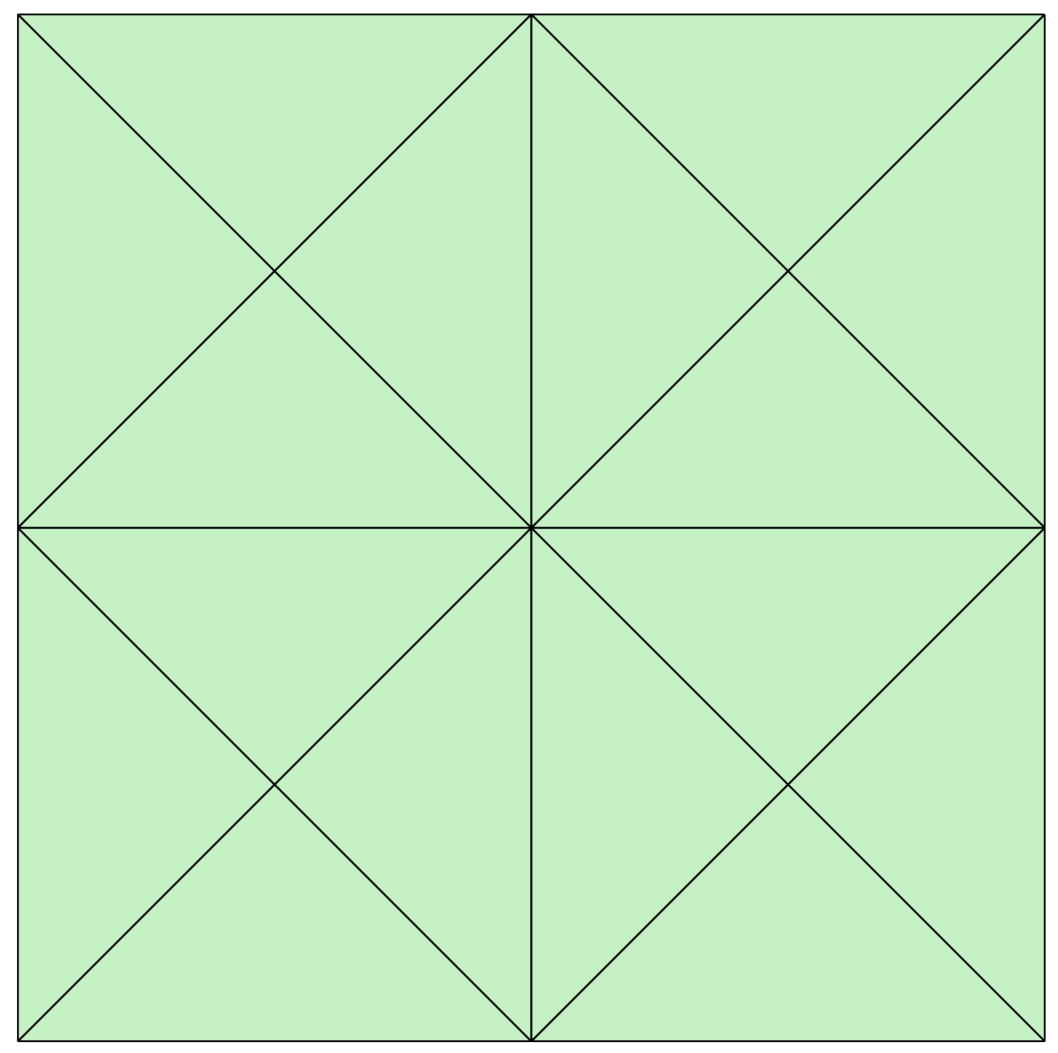}}
	\subfigure[Mesh 2]{\includegraphics[width=0.3\textwidth]{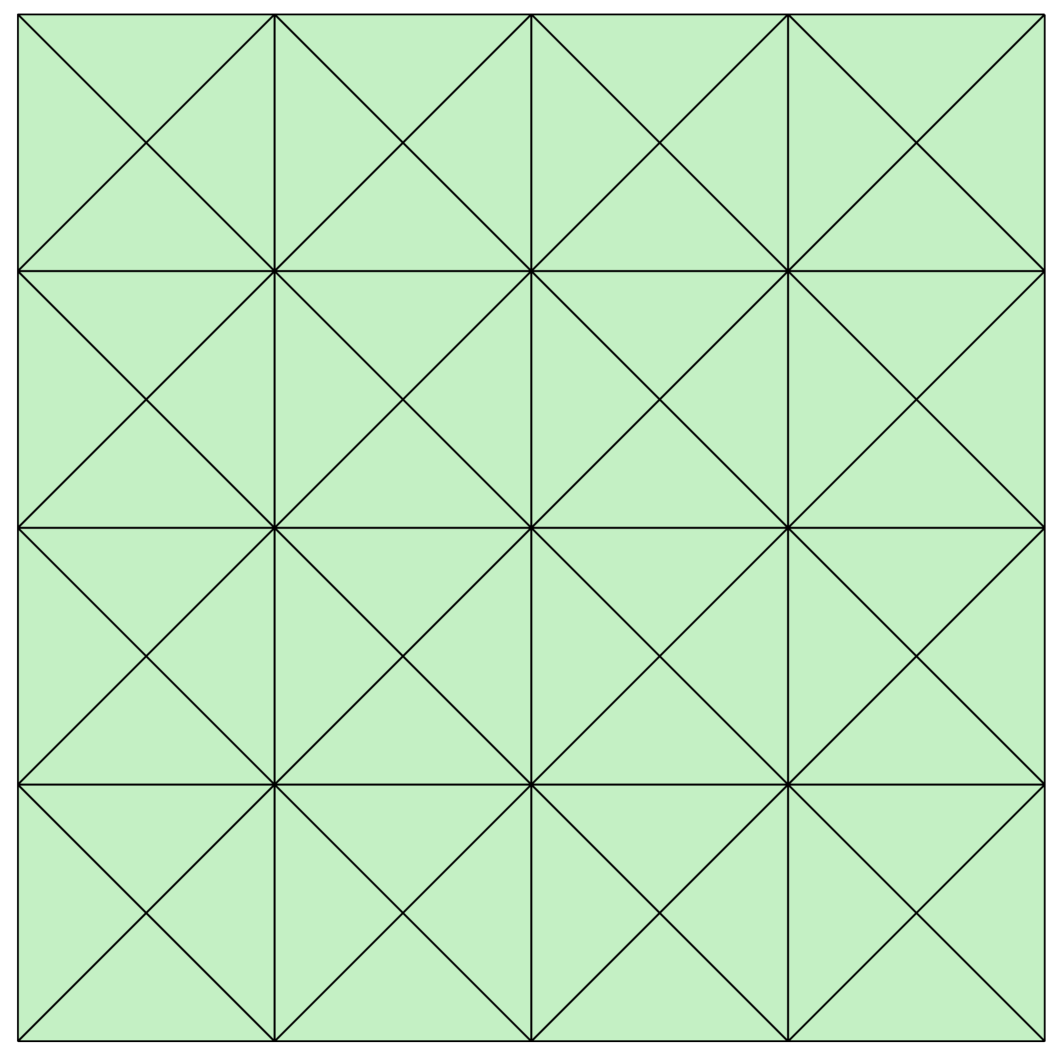}}
	\subfigure[Mesh 3]{\includegraphics[width=0.3\textwidth]{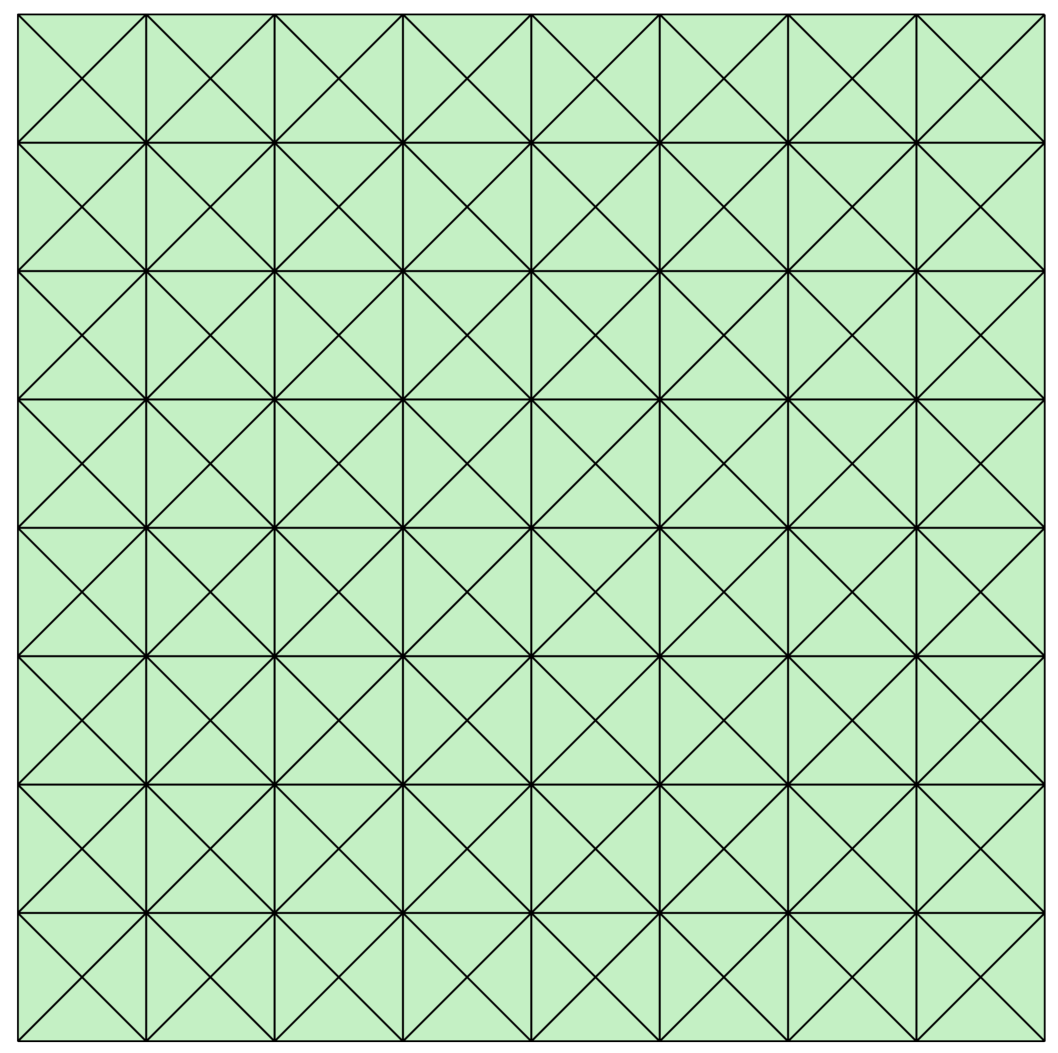}}
	\caption{Triangular meshes of the domain $\Omega = [0,1]^2$ used to test the optimal convergence properties of the HDG method.}
	\label{fig:MeshHDG_2D}
\end{figure}

Four uniform meshes are considered, with 16, 64, 256, and 1,024 triangular elements, respectively. The first three meshes are shown in Figure~\ref{fig:MeshHDG_2D}.

\begin{figure}[!tb]
	\centering
	\subfigure[$\bm{u}$]{\includegraphics[width=0.49\textwidth]{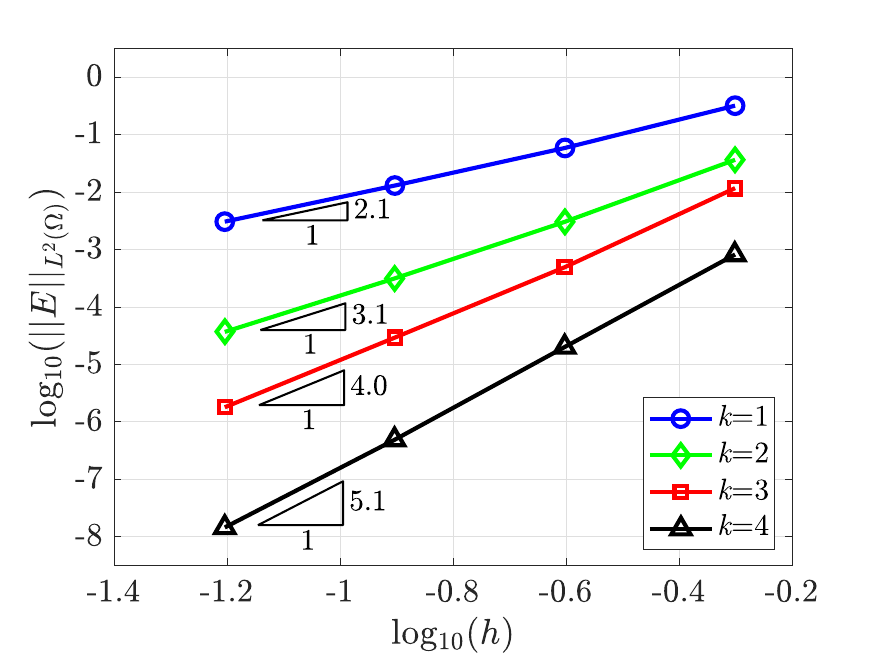}}
	\subfigure[$p$]{\includegraphics[width=0.49\textwidth]{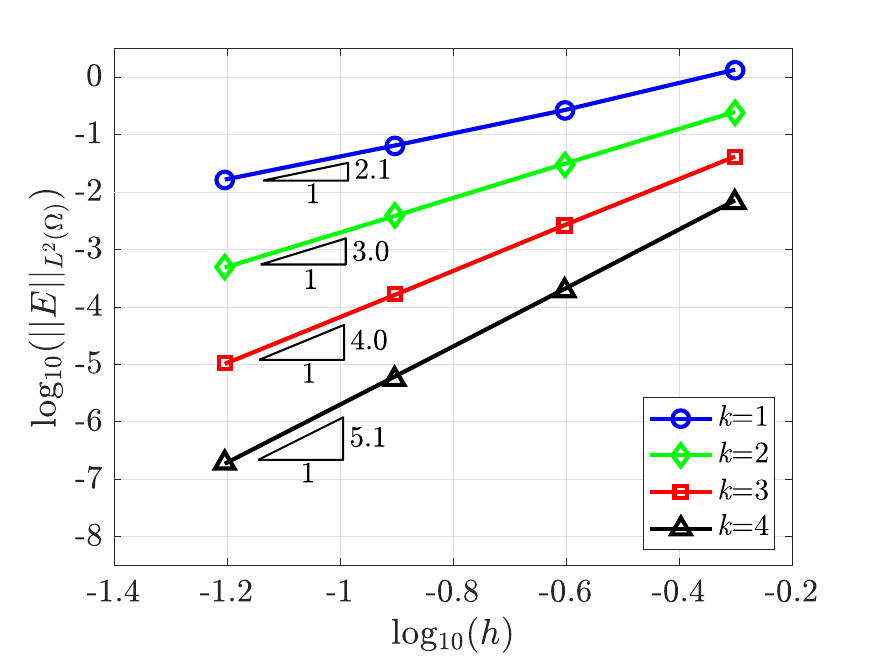}}
	\subfigure[$\bm{L}$]{\includegraphics[width=0.49\textwidth]{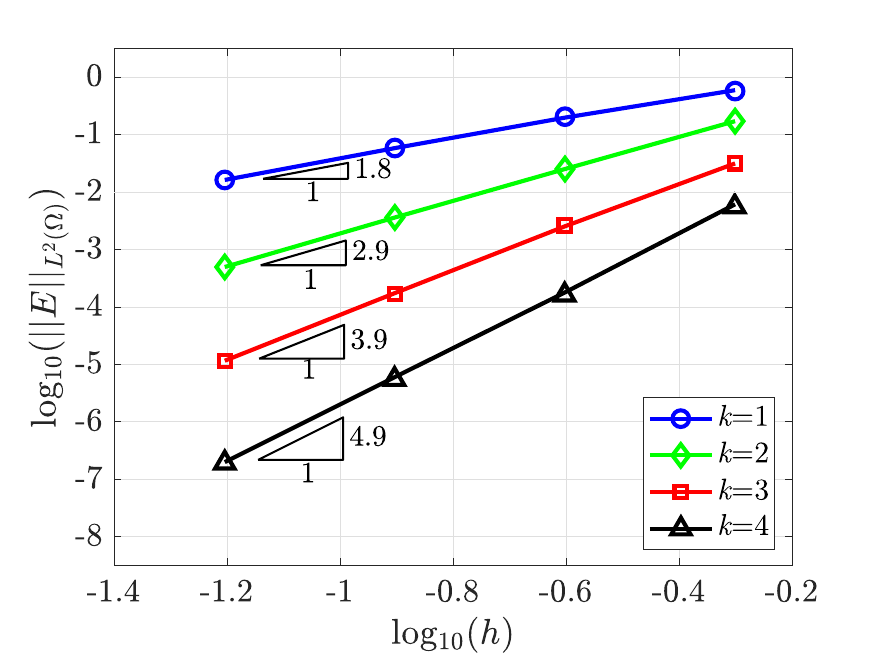}}
	\subfigure[$\bm{u}^\star$]{\includegraphics[width=0.49\textwidth]{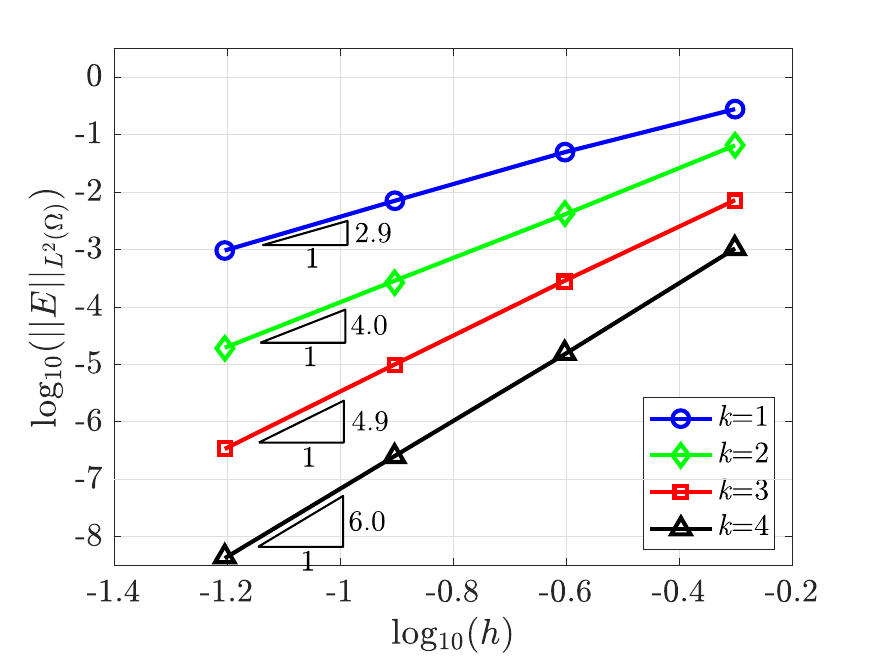}}
	\caption{Kovasznay Flow: $\mathcal{L}_2(\Omega)$ norm of the error for the velocity, $\bm{u}$, pressure, $p$, mixed variable, $\bm{L}$ and postprocessed velocity, $\bm{u}^\star$, as a function of the characteristic element size $h$, for different degrees of approximation.}
	\label{fig:KovasznayhConv}
\end{figure}

Figure~\ref{fig:KovasznayhConv} shows the $\mathcal{L}^2(\Omega)$ norm of the error of the velocity, pressure, velocity gradient and postprocessed velocity as a function of the characteristic element size $h$ for a degree of approximation ranging from $k=1$ up to $k=4$. 
For any degree of approximation, the expected $k+1$ convergence rate can be observed for the velocity, pressure and velocity gradient, whereas the super-convergent velocity shows the $k+2$ convergence rate. The extra accuracy of the super-convergent velocity that allows building an error indicator can be observed.

\subsection{Manufactured transient solution} \label{sc:transientTest}

The second example, considered to verify the correct implementation of the high-order ESDIRK46 time integrator, considers the manufactured solution
\begin{equation}
\bu(\bx) = \begin{Bmatrix}
\sin(x_1 + \omega t) \sin(x_2 + \omega t) \\
\cos(x_1 + \omega t) \cos(x_2 + \omega t)
\end{Bmatrix}, \qquad
p(\bx) = \sin(x_1 - x_2 + \omega t),
\label{eq:manuTime}
\end{equation}
where $\omega = 10$ is used to define a fast variation of all flow quantities in time. The final time used in these examples is $T = 0.25$ and the mesh of Figure~\ref{fig:MeshHDG_2D}(b) is used  with $k=4$ to ensure that the error due to the spatial discretisation is below the error induced by the temporal discretisation. 

\begin{figure}[!tb]
	\centering
	\includegraphics[width=0.49\textwidth]{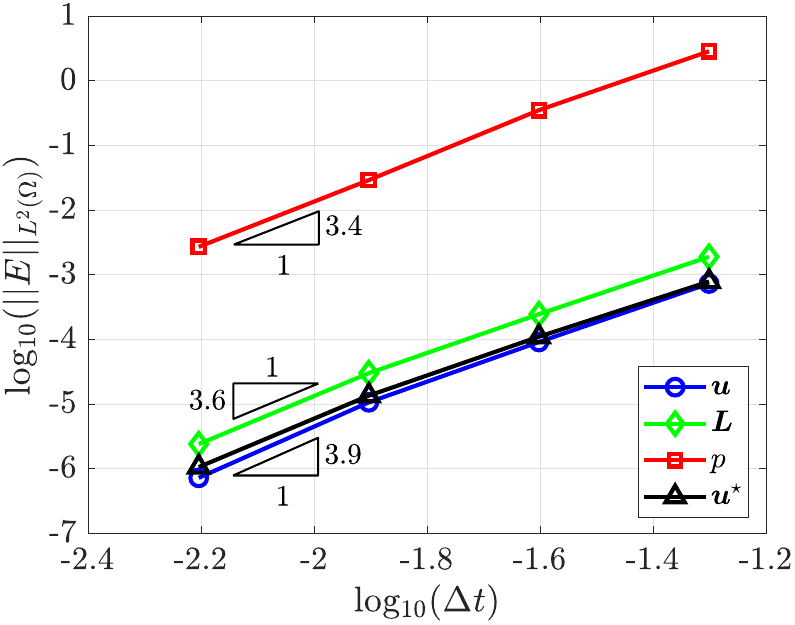}
	\caption{Manufactured transient solution: $\mathcal{L}_2(\Omega)$ norm of the error for the velocity, pressure, velocity gradient and postprocessed velocity, as a function of the time step $\Delta t$.}
	\label{fig:ManufacturedDtConv}
\end{figure}

Figure~\ref{fig:ManufacturedDtConv} shows the $\mathcal{L}^2(\Omega)$ norm of the error for the velocity, pressure, velocity gradient and postprocessed velocity as a function of the time step $\Delta t$.

The observed convergence rates generally align with the theoretical fourth order of accuracy for the ESDIRK46 method. The slightly lower rate observed for the pressure is associated to the so-called order reduction of ESDIRK methods~\cite{sanz1986convergence} often observed when non-homogeneous boundary conditions are considered.

\subsection{Flow around two circular cylinders} \label{sc:twoCyls}

The next example considers the laminar flow, at $Re=100$, around two circular cylinders on tandem. The far field is made of a circle of diameter 100 centred at the origin, whereas the two circular cylinders have diameter 1 and are centred at $(-20,0)$ and $(10,0)$, respectively. 

\begin{figure}[!tb]
	\centering
	\includegraphics[width=\textwidth]{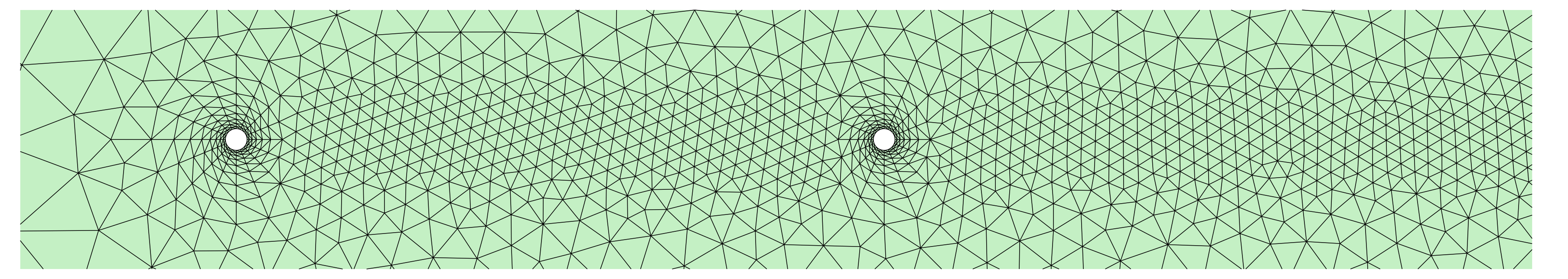}
	\caption{Flow around two circular cylinders: detail of the unstructured triangular mesh near the circular cylinders.}
	\label{fig:twoCyls_Mesh}
\end{figure}

An unstructured mesh of 2,712 triangles is employed for this example. Curved elements are generated near the cylinder using the elastic analogy presented in~\cite{xie2013generation}. Given the low Reynolds number considered, the size of the elements in the normal direction to the wall is relatively large and only the first two layers of elements around the cylinders are curved. More precisely, the size of the first element around the circular cylinders is 0.01 and the growing factor in the normal direction is 1.4. Two point sources are introduced to prescribe a mesh size of 0.2 near the cylinders, whereas a line source with size 0.75 is placed in the path of the von Karman vortex street. A detailed view of the mesh near the cylinders is shown in Figure~\ref{fig:twoCyls_Mesh}.

The ESDIRK46 time marching algorithm~\cite{kennedy2016diagonally} is used with a time step $\Delta t = 0.2$ and the solution is advanced until the final time $T=200$.

\begin{figure}[!tb]
	\centering
	\subfigure[Pressure]{\includegraphics[width=\textwidth]{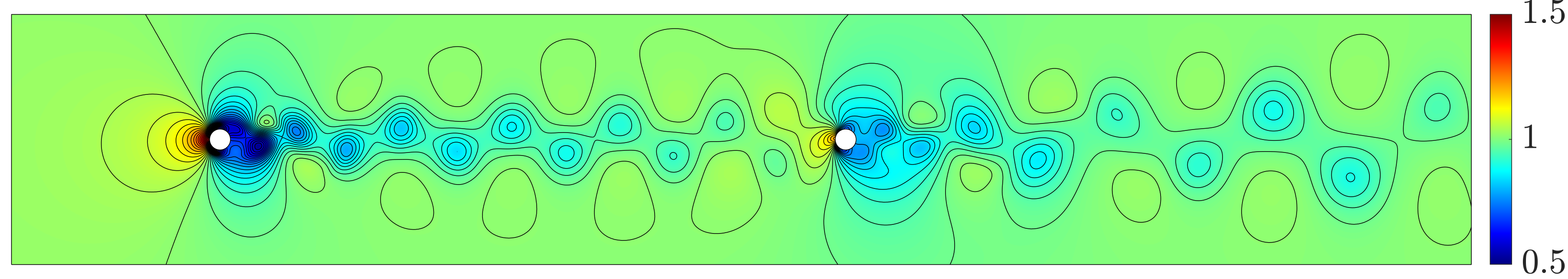}}
	\subfigure[Velocity]{\includegraphics[width=\textwidth]{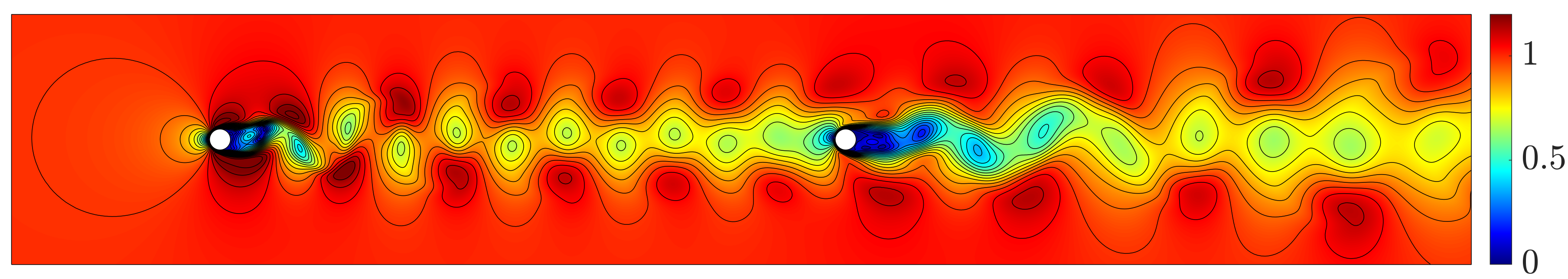}}
	\caption{Flow around two circular cylinders: Pressure and magnitude of the velocity fields at $t=200$ with a uniform degree of approximation $k=6$.}
	\label{fig:twoCyls_Ref}
\end{figure}

As no analytical solution is available for this problem, a reference solution is computed by employing a uniform degree of approximation $k=6$. Further numerical experiments, not reported here for brevity, were performed to ensure that $k=6$ is the minimum degree that is required for this problem to get a converged solution. Figure~\ref{fig:twoCyls_Ref} shows the reference pressure and magnitude of the velocity at $t=200$.

\begin{figure}[!tb]
	\centering
	\subfigure[Pressure]{\includegraphics[width=\textwidth]{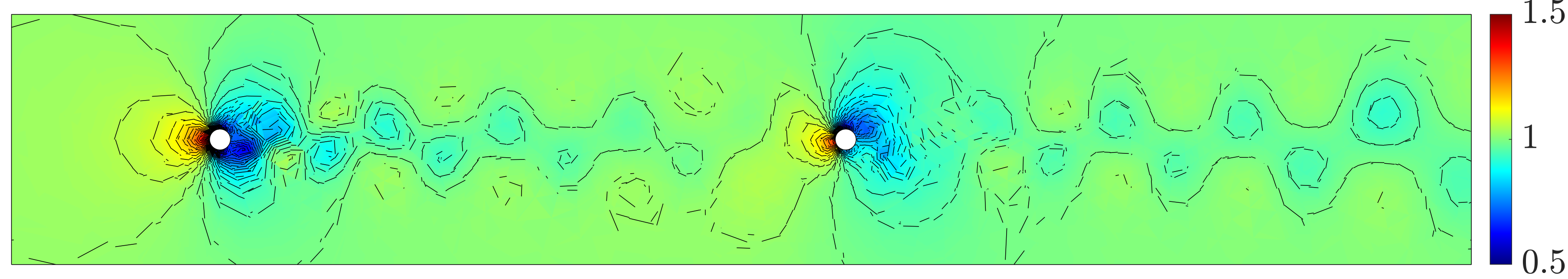}}
	\subfigure[Velocity]{\includegraphics[width=\textwidth]{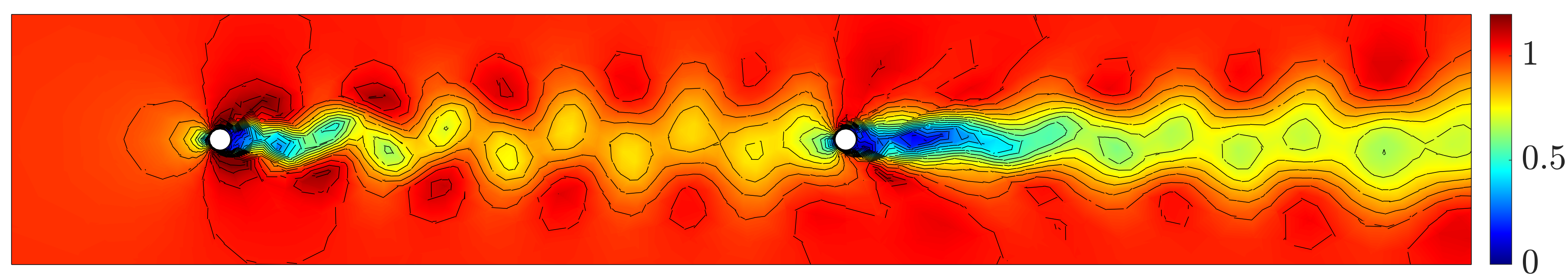}}
	\caption{Flow around two circular cylinders: Pressure and magnitude of the velocity fields at $t=200$ with a uniform degree of approximation $k=1$.}
	\label{fig:twoCyls_P1}
\end{figure}

The high-order spatial approximation is crucial in this problem to accurate capture the von Karman vortex street generated by the first cylinder and its influence on the second cylinder. If a low order ($k=1$) approximation is used on the same mesh, the intensity of the vortices is not captured, as shown in Figure~\ref{fig:twoCyls_P1}, clearly illustrating the low dissipative properties of a high-order approximation scheme.
The low order results also display a larger dispersion when compared to the high order approximation as the vortices appear in different positions.

The results shown in Figure~\ref{fig:twoCyls_Ref} suggest that a uniform degree of approximation is not required and a degree adaptive approach is an attractive approach to increase the resolution only where is needed. The next experiments compare different degree adaptive strategies where the desired error, as detailed in Section~\eqref{sc:error}, is taken as $\varepsilon = 10^{-4}$ in all experiments, unless otherwise stated. Given the large time step used, the adaptive process is repeated twice per time step to ensure that flow features are properly captured as the solution progresses. The effect of not repeating the adaptive process is also illustrated in this example.

\begin{figure}[!tb]
	\centering
	\subfigure[Degree map]{\includegraphics[width=\textwidth]{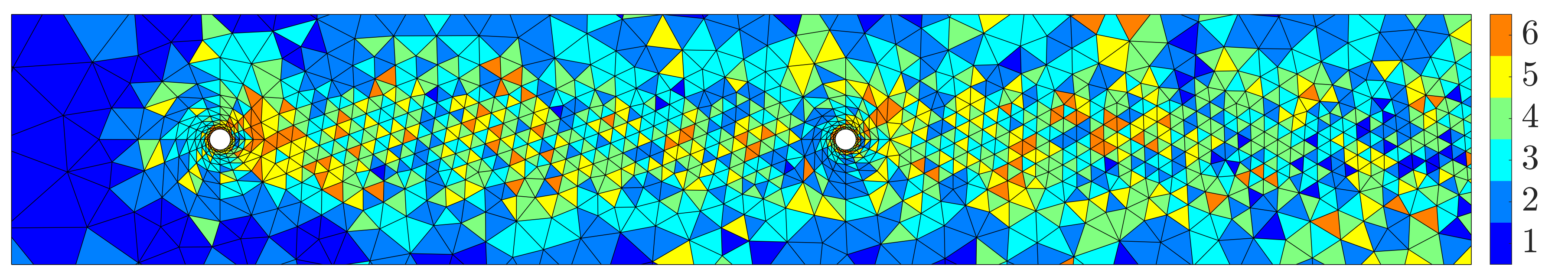}}
	\subfigure[Pressure]{\includegraphics[width=\textwidth]{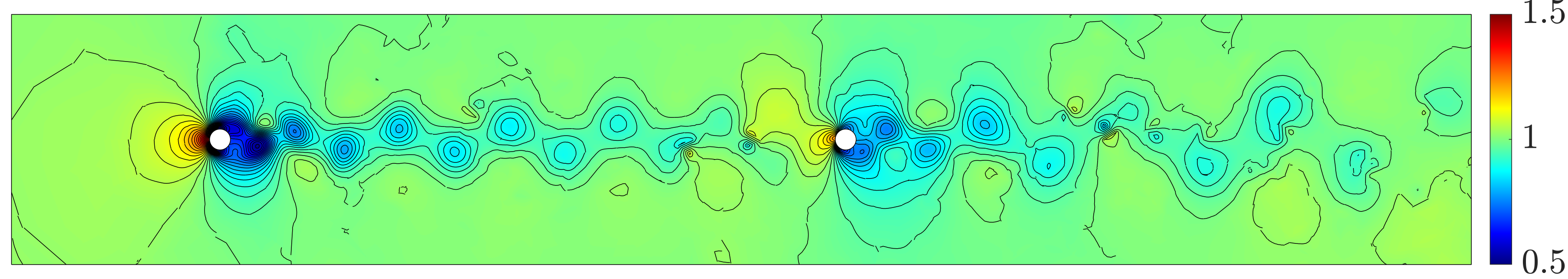}}
	\subfigure[Velocity]{\includegraphics[width=\textwidth]{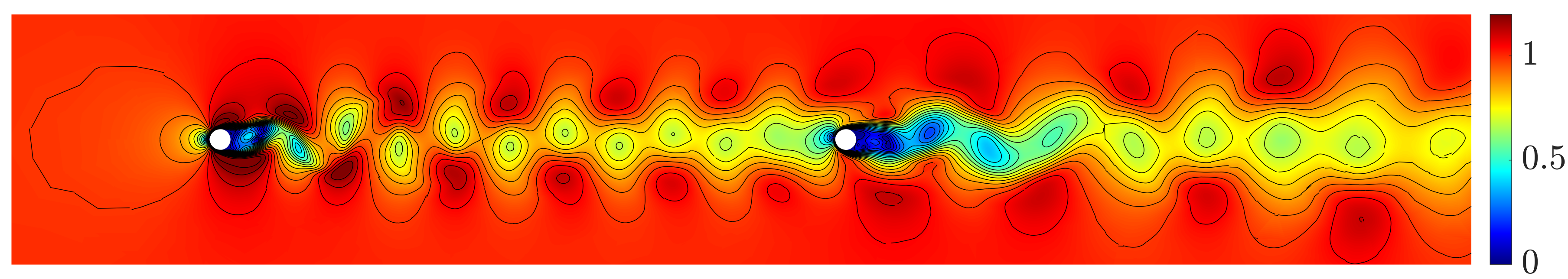}}
	\caption{Flow around two circular cylinders: Pressure and magnitude of the velocity fields at $t=200$ with degree adaptivity.}
	\label{fig:twoCyls_Adap}
\end{figure}

A standard degree adaptive approach, i.e. without the proposed correction, is first considered, where at each time step the solution in each element is projected using the desired degree of approximation map according to the error indicator provided by the HDG method. The results at $t=200$ are shown in Figure~\ref{fig:twoCyls_Adap}, including the degree used in each element. 
The velocity field is in good agreement with the reference solution, with only a minor loss of intensity of the vortices behind the circular cylinders. However, the pressure field shows some important numerical artefacts when compared to the reference solution, which are related to the violation of the incompressibility constraint when projecting the velocity field from a given degree map to another degree map, in particular when the degree of approximation is decreased, as explained in Section~\ref{sc:conservativeProj}.

\begin{figure}[!tb]
	\centering
	\subfigure[Lift]{\includegraphics[width=0.48\textwidth]{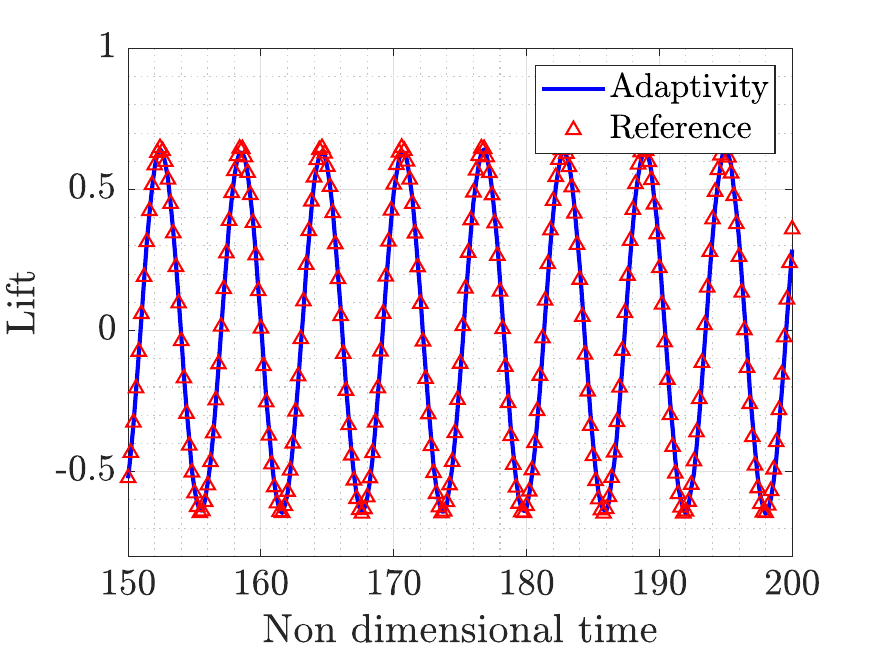}}
	\subfigure[Drag]{\includegraphics[width=0.48\textwidth]{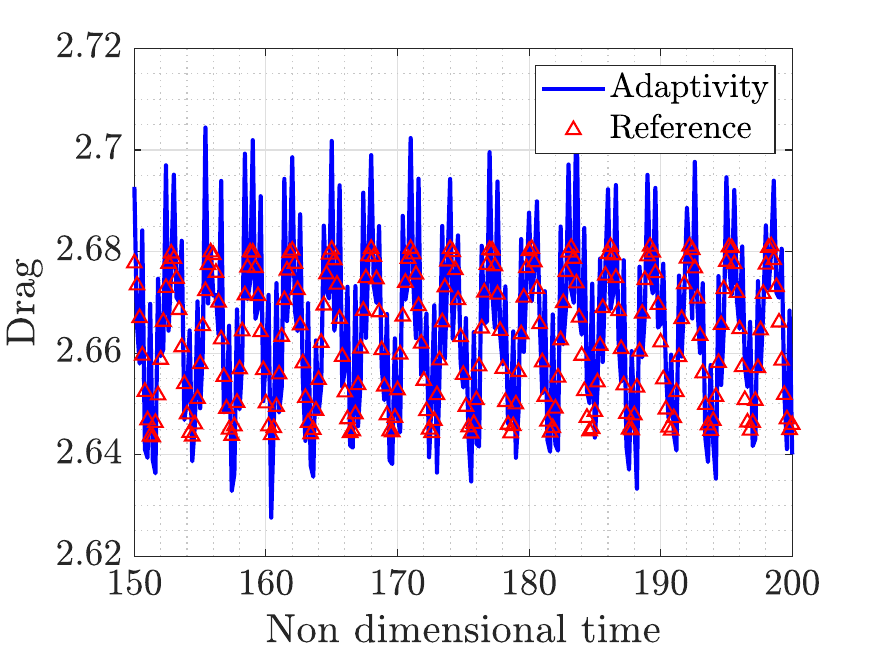}}
	\caption{Flow around two circular cylinders: lift and drag over the first cylinder using degree adaptivity compared to the reference solution.}
	\label{fig:TwoCyl_Obj1NoCorrect}
\end{figure}

\begin{figure}[!tb]
	\centering
	\subfigure[Lift]{\includegraphics[width=0.48\textwidth]{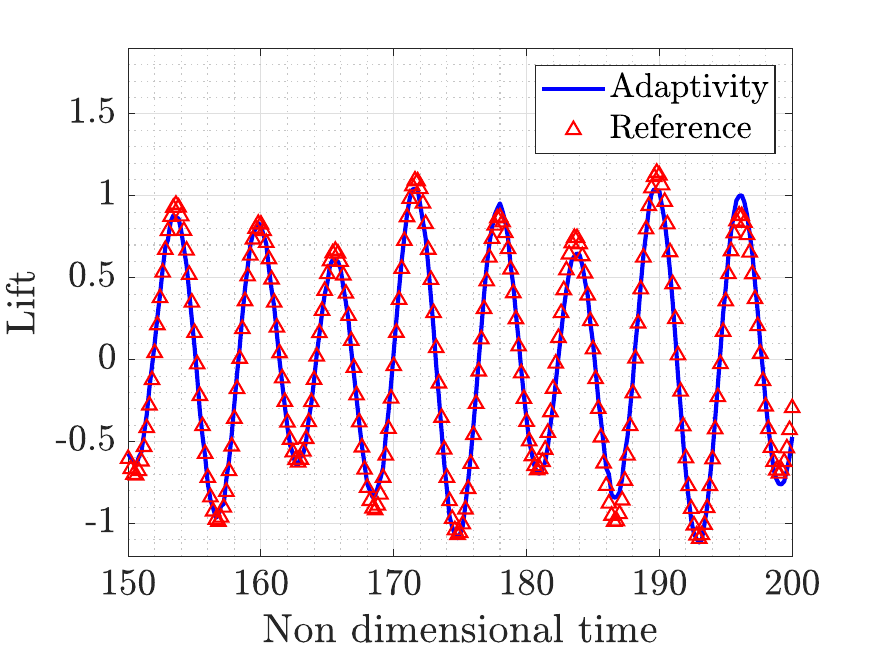}}
	\subfigure[Drag]{\includegraphics[width=0.48\textwidth]{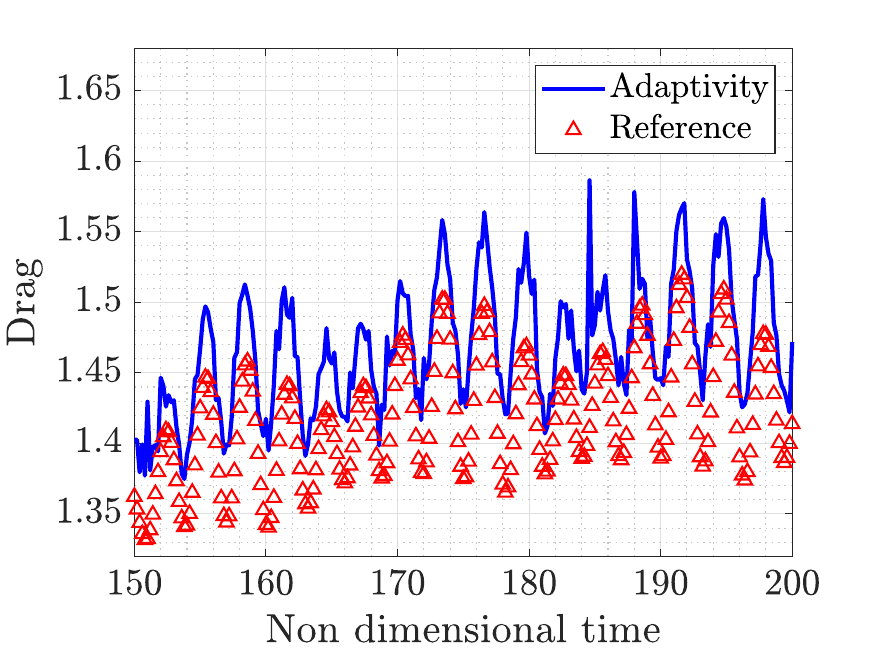}}
	\caption{Flow around two circular cylinders: lift and drag over the second cylinder using degree adaptivity compared to the reference solution.}
	\label{fig:TwoCyl_Obj2NoCorrect}
\end{figure}

To quantify the accuracy of the simulations, the lift and drag are considered the  quantities of interest. Figure~\ref{fig:TwoCyl_Obj1NoCorrect} shows the lift and drag on the first cylinder using a standard degree adaptive approach, i.e. without the proposed correction, and the result is compared to the reference solution.
The results clearly display non-physical oscillations of the drag, whereas the lift is accurately computed. Similar results for the quantities of interest for the second cylinder are shown in Figure~\ref{fig:TwoCyl_Obj2NoCorrect}.

Further experiments have been performed to confirm that the apparent more accurate results on the lift are due to the cancellation of errors and the symmetry of the lift with respect to a zero mean value. To corroborate this a mesh convergence analysis has been performed for the steady flow around a cylinder at $Re=30$, measuring the lift and drag on the upper and lower parts of the cylinders. The results show that the values of the lift and drag, measured separately on the upper and lower parts of the cylinder, converge to reference values as the mesh is refined. However, when the error of the total lift and total drag are measured, only the drag shows the expected reduction of the error as the mesh is refined, whereas the lift exhibits a very small error, even on coarse meshes, due to the addition of the upper and lower contributions, which have opposite sign.

\begin{figure}[!tb]
	\centering
	\subfigure[Degree map]{\includegraphics[width=\textwidth]{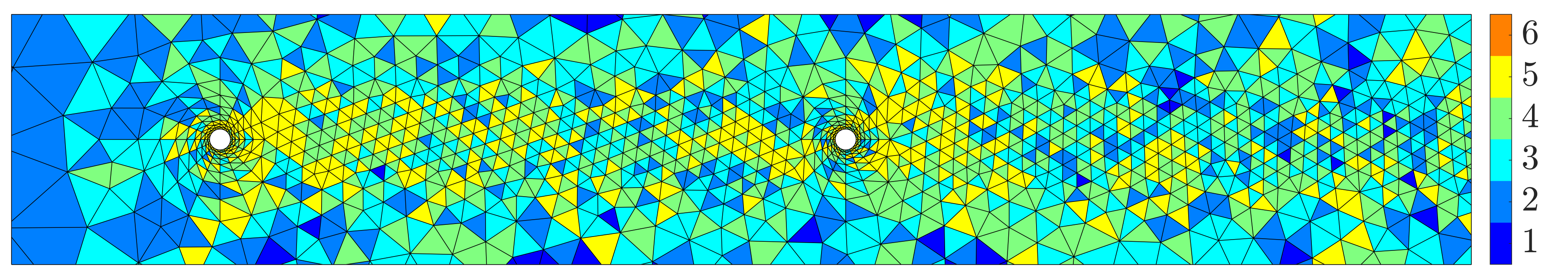}}
	\subfigure[Pressure]{\includegraphics[width=\textwidth]{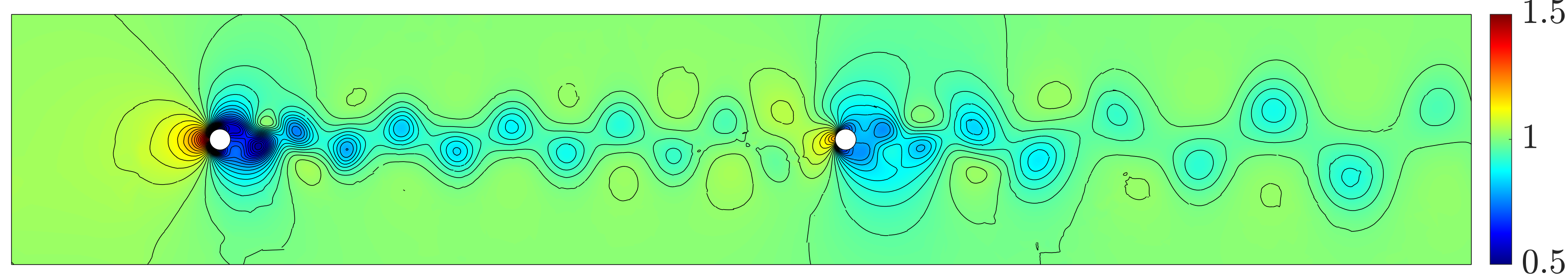}}
	\subfigure[Velocity]{\includegraphics[width=\textwidth]{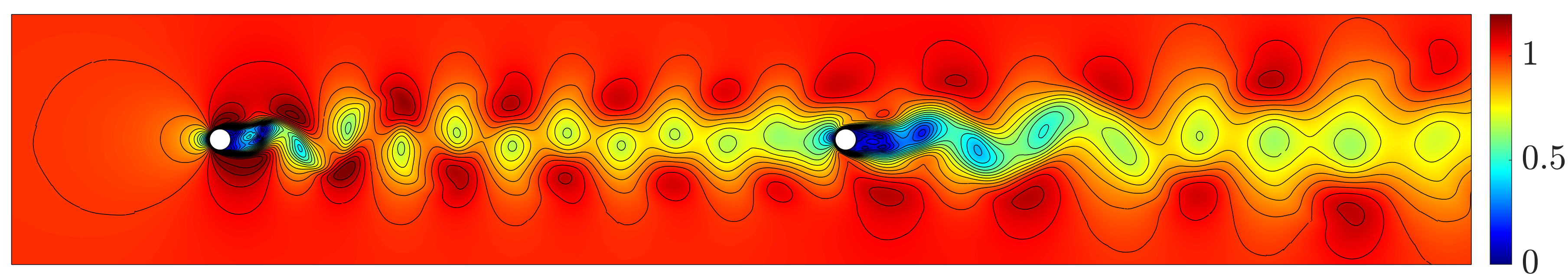}}
	\caption{Flow around two circular cylinders: Pressure and magnitude of the velocity fields at $t=200$ with degree adaptivity and the conservative projection.}
	\label{fig:twoCyls_AdapCorrect}
\end{figure}

Next, the degree adaptive procedure is enhanced by introducing the correction proposed in Section~\ref{sc:conservativeProj}. To illustrate the benefits of the proposed approach, Figure~\ref{fig:twoCyls_AdapCorrect} shows the degree map, pressure and magnitude of the velocity at $t=200$.
It can be observed that all the artefacts on the pressure field are not present and an excellent agreement with the reference solution is obtained.

\begin{figure}[!tb]
	\centering
	\subfigure[Lift]{\includegraphics[width=0.48\textwidth]{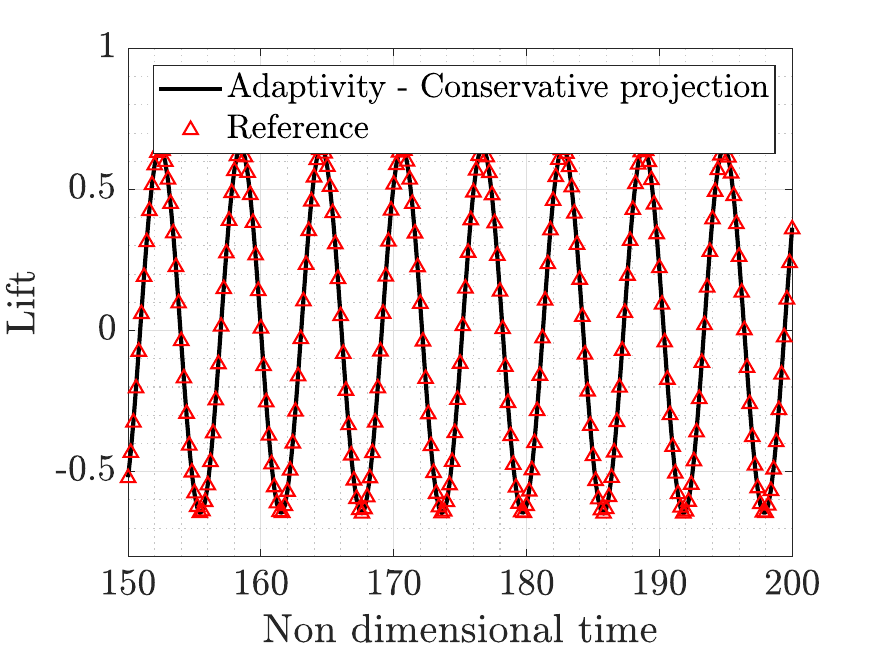}}
	\subfigure[Drag]{\includegraphics[width=0.48\textwidth]{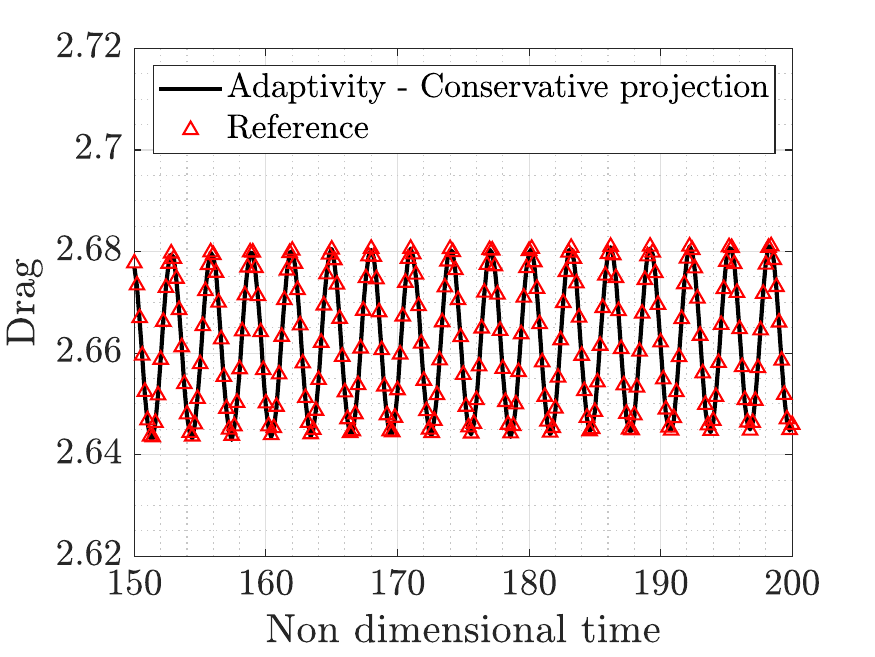}}
	\caption{Flow around two circular cylinders: lift and drag over the first cylinder using degree adaptivity and the proposed correction compared to the reference solution.}
	\label{fig:TwoCyl_Obj1Correct}
\end{figure}

\begin{figure}[!tb]
	\centering
	\subfigure[Lift]{\includegraphics[width=0.48\textwidth]{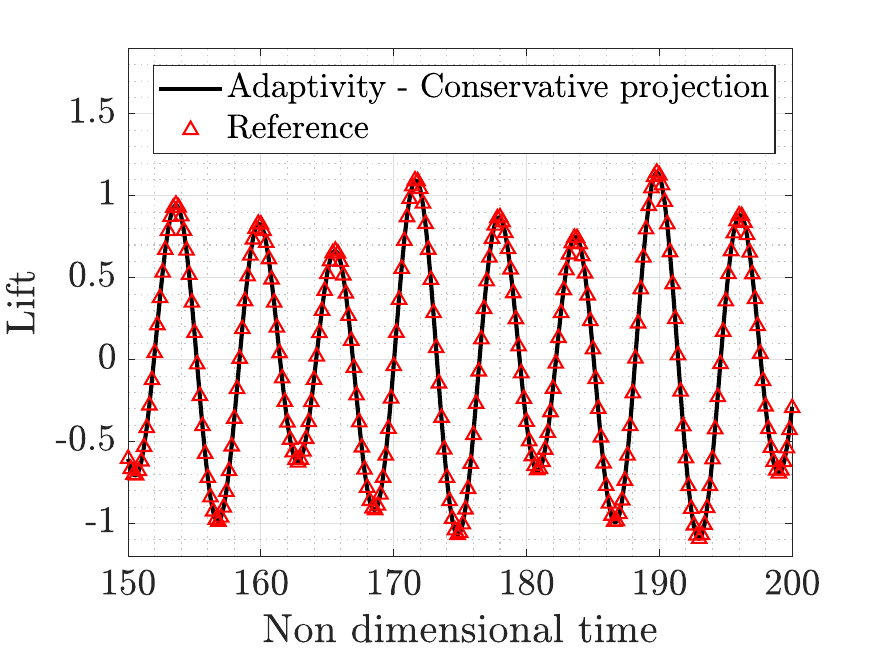}}
	\subfigure[Drag]{\includegraphics[width=0.48\textwidth]{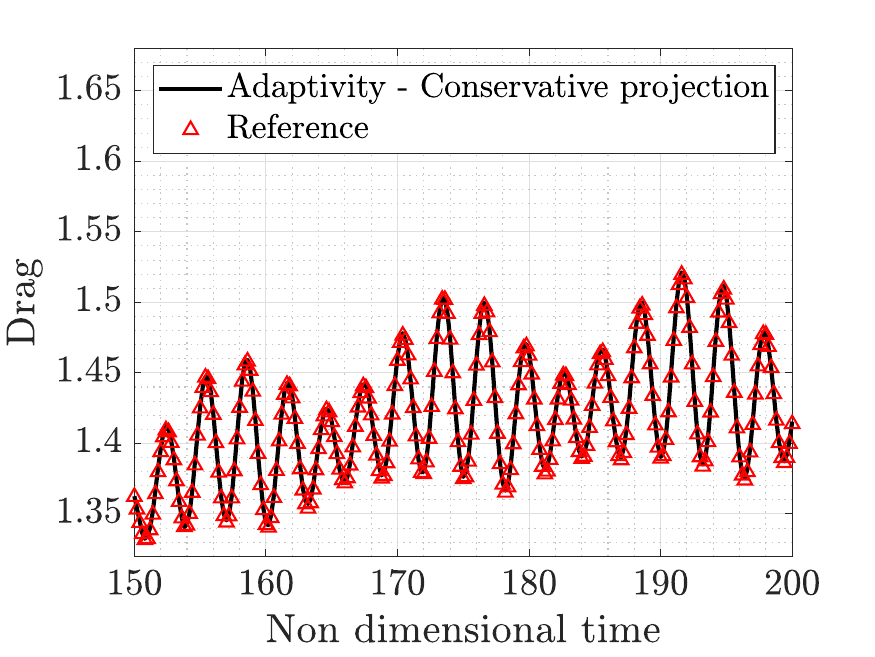}}
	\caption{Flow around two circular cylinders: lift and drag over the second cylinder using degree adaptivity and the proposed correction compared to the reference solution.}
	\label{fig:TwoCyl_Obj2Correct}
\end{figure}

To better quantify the accuracy of the simulation with the proposed conservative projection, Figure~\ref{fig:TwoCyl_Obj1Correct} shows the lift and drag on the first cylinder.
The results demonstrate that the proposed correction completely removes the non-physical oscillations shown in the previous simulations and provide a lift and drag which are in excellent agreement with the reference solution. The results for the second cylinder are shown in Figure~\ref{fig:TwoCyl_Obj2Correct}, showing again that no oscillations are observed and a very good agreement with the reference solution is obtained.

To further illustrate the benefit of the proposed conservative projection, Table~\ref{tb:twoCylsLiftDragErr} reports the maximum error of the lift and drag for both cylinders.
\begin{table}[!tb]
	\centering
	\begin{tabular}{|c|c|c|c|c|}
		\hline
		& \multicolumn{2}{|c|}{Cylinder 1} & \multicolumn{2}{|c|}{Cylinder 2} \\  
		\hline
		& Standard & Conservative  & Standard & Conservative \\
		& adaptivity & projection  & adaptivity & projection\\
		\hline
		Lift error & $8.1 \times 10^{-2}$ & $6.8 \times 10^{-3}$ & $1.8 \times 10^{-1}$ & $1.5 \times 10^{-2}$ \\ 
		Drag error & $3.7 \times 10^{-2}$ & $1.0 \times 10^{-3}$ & $1.8 \times 10^{-1}$ & $4.6 \times 10^{-3}$ \\
		\hline
	\end{tabular}
	\caption{Flow around two circular cylinders: maximum error in lift and drag for the two cylinders using the standard adaptivity and the adaptivity with the proposed conservative projection.}
	\label{tb:twoCylsLiftDragErr}
\end{table}
The results clearly show the extra accuracy provided by the conservative projection. More precisely the error in the lift is more than 10 times lower using the conservative projection, whereas the error in the drag is almost 40 times lower.

To conclude this example, further numerical experiments are performed to illustrate that the conservative projection is only needed when the degree of approximation is allowed to decrease during the adaptive process. In addition, the effect of the desired error during the degree adaptive process is illustrated.

\begin{figure}[!tb]
	\centering
	\subfigure[Object 1]{\includegraphics[width=0.48\textwidth]{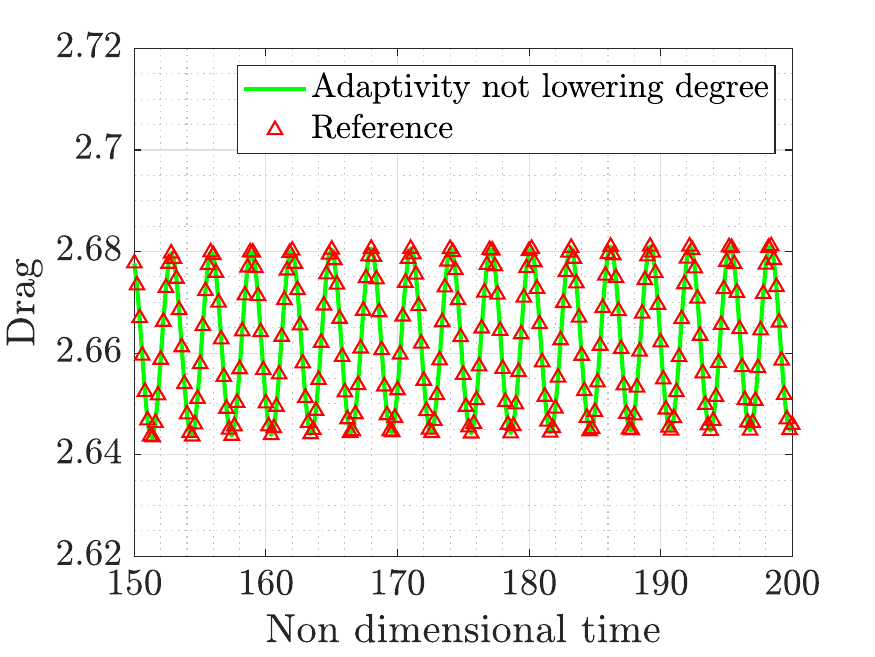}}
	\subfigure[Object 2]{\includegraphics[width=0.48\textwidth]{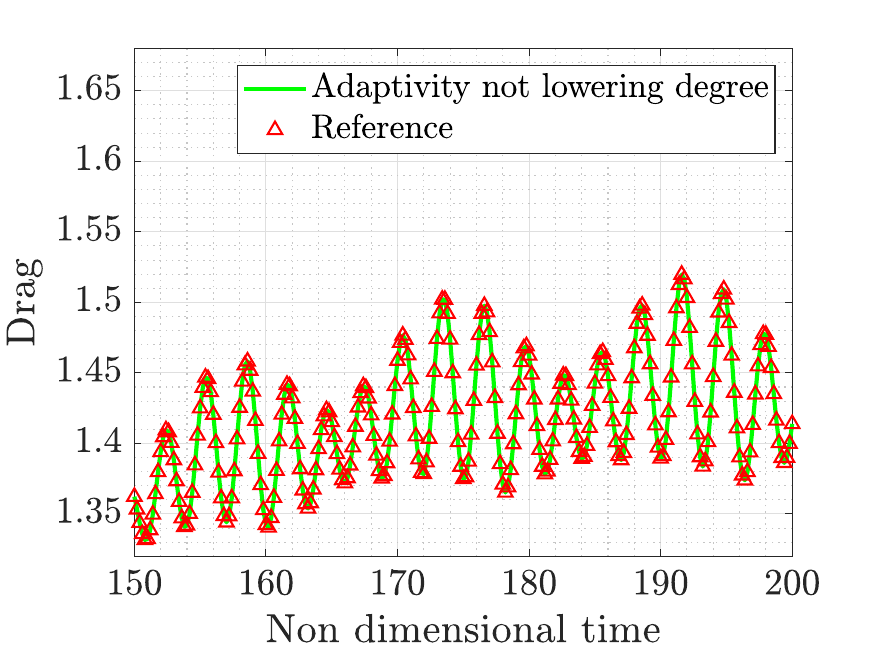}}
	\caption{Flow around two circular cylinders: Drag on the two cylinders using degree adaptivity and not allowing the degree to be decreased during the adaptive process.}
	\label{fig:TwoCyl_DragNoLowering1EM4}
\end{figure}

Figure~\ref{fig:TwoCyl_DragNoLowering1EM4} shows the drag on the first and second cylinders using a standard degree adaptivity where the degree of approximation is not allowed to decrease.
It can be observed that a very good agreement with the reference solution is obtained, without the oscillatory behaviour that was observed when the degree was allowed to decrease during the adaptive process. However, the main drawback of this approach is the obvious increase of computational cost because if an element reaches a high degree of approximation at one time step, the degree will be maintained at such degree for the rest of the simulation, even if there is no need to capture any features at that region for the remaining of the simulation. In this example, due to the impulsive start and the low desired error in each element $\varepsilon = 10^{-4}$, all elements of the mesh require, at some instant, a degree of approximation $k=6$, so this approach leads to the same solution as the reference solution with the extra cost of computing the error indicator  and projecting the solution at each time step. 

\begin{figure}[!tb]
	\centering
	\subfigure[Object 1]{\includegraphics[width=0.48\textwidth]{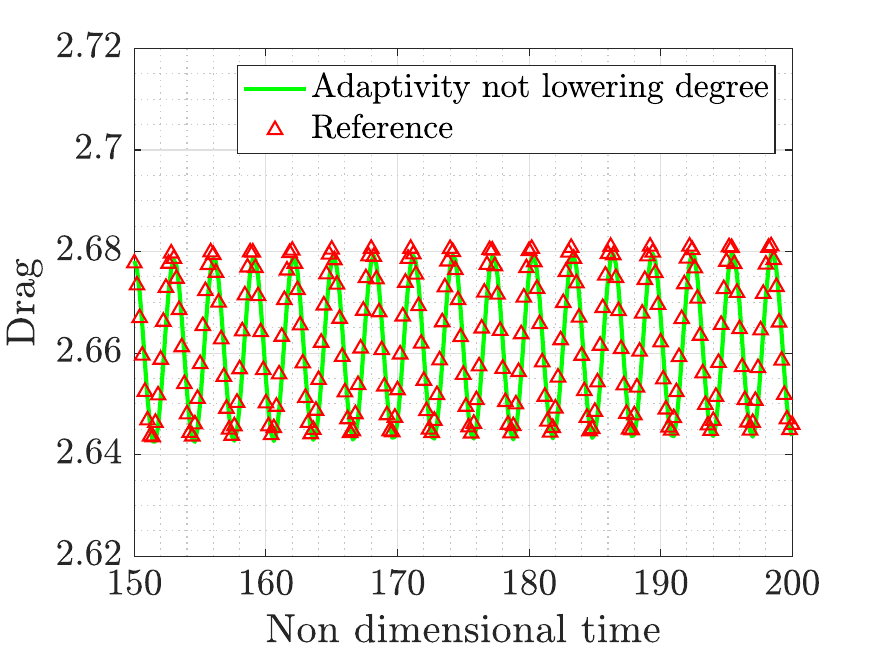}}
	\subfigure[Object 2]{\includegraphics[width=0.48\textwidth]{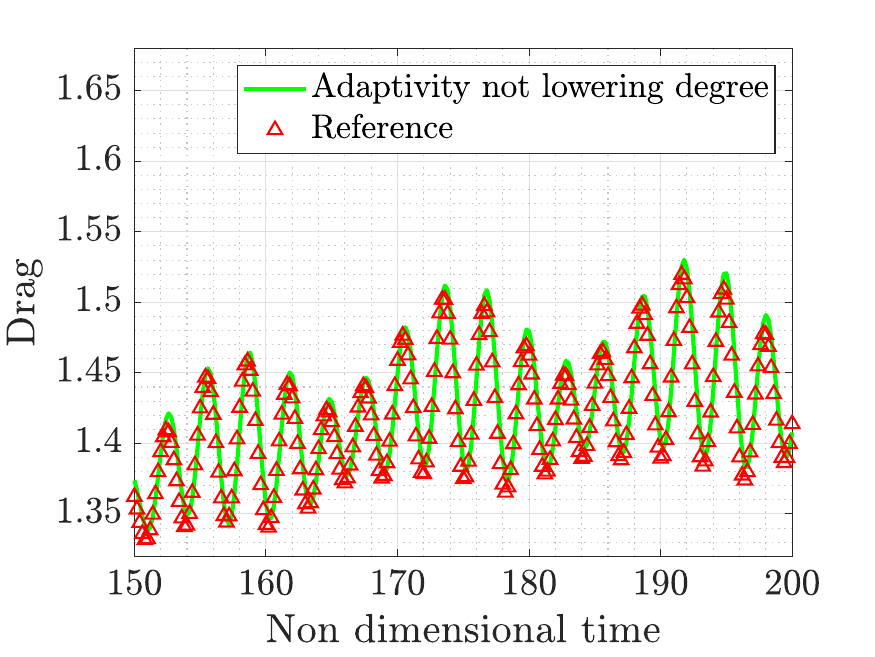}}
	\caption{Flow around two circular cylinders: Drag on the two cylinders using degree adaptivity and not allowing the degree to be decreased during the adaptive process with $\varepsilon = 10^{-3}$.}
	\label{fig:TwoCyl_DragNoLowering1EM3}
\end{figure}

If a less restrictive tolerance is used in the adaptive process, namely $\varepsilon = 10^{-3}$, the quantities of interest are obtained without oscillations, as shown in Figure~\ref{fig:TwoCyl_DragNoLowering1EM3}, providing evidence that the cause for the oscillations in the drag is the violation of the incompressibility condition during the projection of the solution to a lower degree.
Some discrepancies in the drag of the second cylinder are visually observed due to the use of a less restrictive tolerance.

\begin{figure}[!tb]
	\centering
	\includegraphics[width=\textwidth]{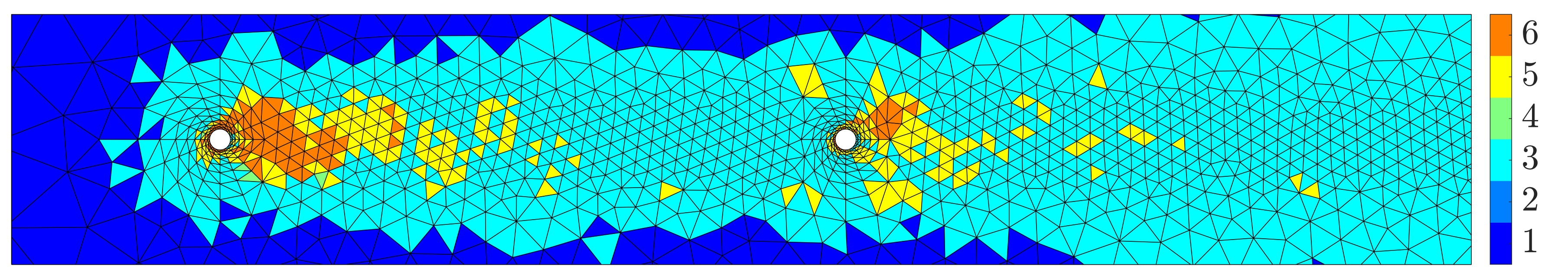}
	\caption{Flow around two circular cylinders: Degree of approximation at $t=200$ not allowing the degree to be decreased during the adaptive process.}
	\label{fig:twoCyls_NotLoweringP}
\end{figure}

The degree map at $t=200$ when the adaptive process is implemented without allowing the degree of approximation to be decreased and with $\varepsilon = 10^{-3}$ is shown in Figure~\ref{fig:twoCyls_NotLoweringP}. 
Compared to the degree map of the adaptivity process with the proposed correction, shown in Figure~\ref{fig:twoCyls_AdapCorrect}, it can be observed that the majority of elements in the wake of the two cylinders is kept to a higher degree when the adaptivity process is not allowed to lower the degree. It is also noticeable that when the degree is not allowed to decrease, a number of elements in the wake of the two cylinders end up using a degree of approximation $k=6$ whereas if the adaptivity is allowed to decrease the degree, this high degree of approximation is not required at the final time.

\begin{figure}[!tb]
	\centering
	\includegraphics[width=0.48\textwidth]{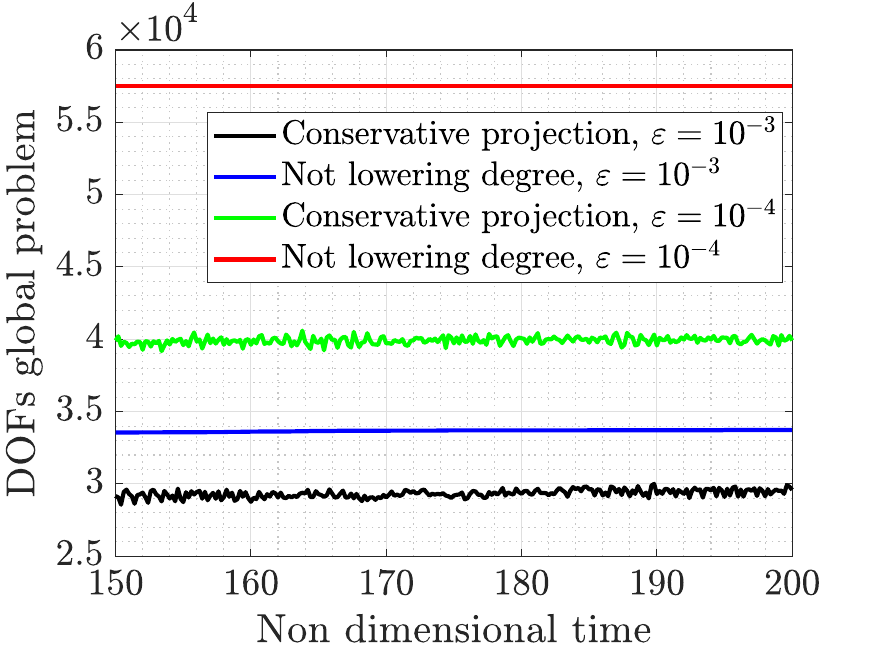}
	\caption{Flow around two circular cylinders: Number of degrees of freedom of the global problem for two different adaptive approaches and for two different values of the desired error.}
	\label{fig:TwoCyl_dofComparison}
\end{figure}

To quantify the reduction of degrees of freedom induced by allowing the adaptive process to decrease the degree of approximation is shown in Figure~\ref{fig:TwoCyl_dofComparison}
The results clearly illustrate the advantage of using the proposed projection to enable the adaptive process to lower the degree during the time marching procedure. It is also worth noting that the lower the desired error, the more advantageous is to allow the degree to be lowered.

In terms of computational cost, the simulation with the proposed conservative projection is almost two times faster than the simulation with a uniform degree of approximation $k=6$. The simulation with the  conservative projection is more than three times faster than the simulation not lowering the degree. The simulation not lowering the degree is actually more expensive than computing the reference solution because the majority of elements end up having the maximum degree of approximation but the cost of computing the error indicator and projecting the solution twice every time step becomes important.  This shows that the reduction in degrees of freedom translates in an important reduction in computational time.

\begin{figure}[!tb]
	\centering
	\subfigure[Object 1]{\includegraphics[width=0.48\textwidth]{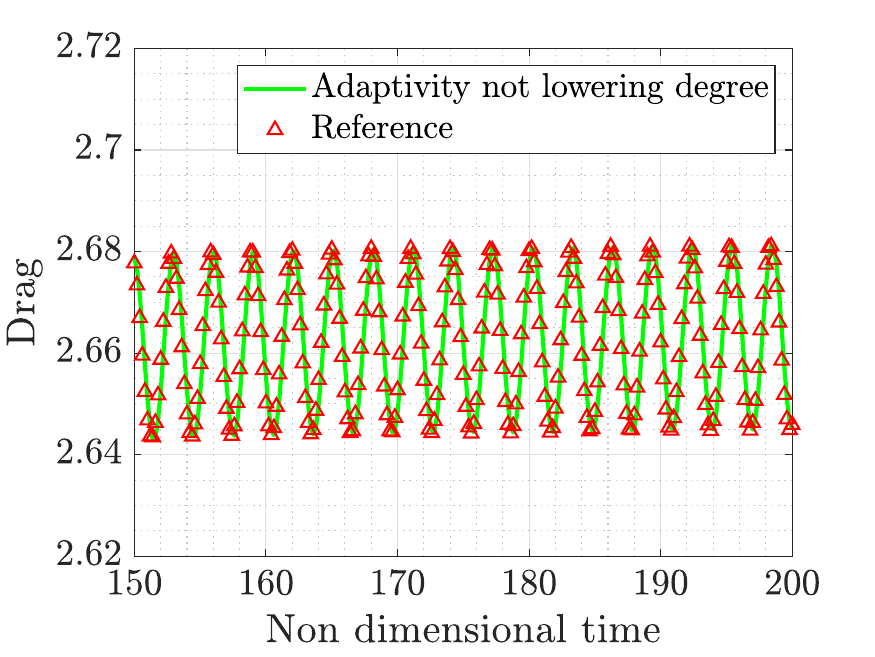}}
	\subfigure[Object 2]{\includegraphics[width=0.48\textwidth]{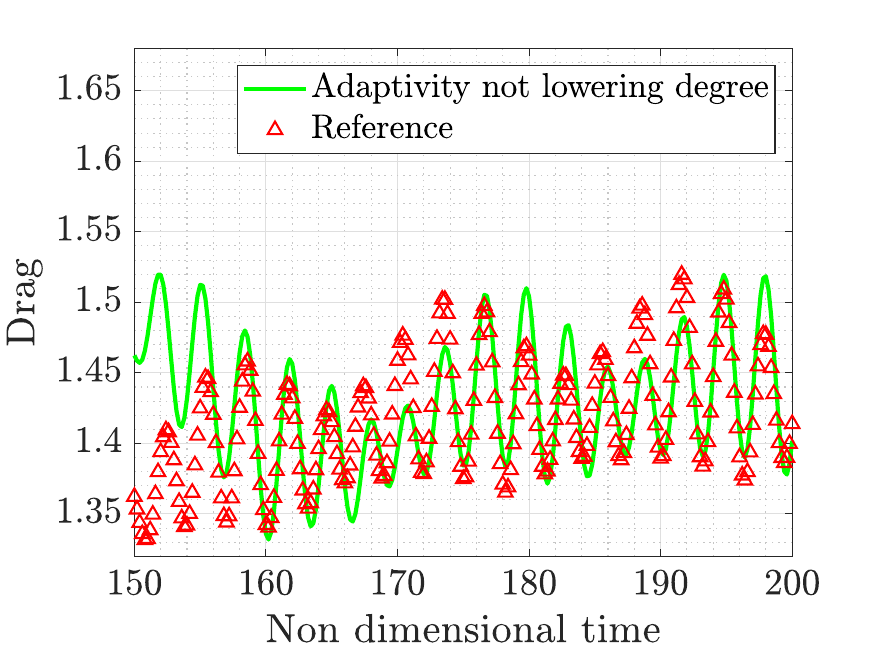}}
	\caption{Flow around two circular cylinders: Drag on the two cylinders using degree adaptivity, not allowing the degree to be decreased during the adaptive process with $\varepsilon = 10^{-3}$ and performing the adaptivity only once per time step.}
	\label{fig:TwoCyl_DragNoRepeating1EM3}
\end{figure}

Finally, the need to repeat the adaptive process twice at each time step is also illustrated using a numerical experiment. The simulation of Figure~\ref{fig:TwoCyl_DragNoLowering1EM3} is repeated but performing the degree adaptivity only once per time step. Due to the large time step used with a high order time integrator, the computed drag shows a significant loss of accuracy, as shown in Figure~\ref{fig:TwoCyl_DragNoRepeating1EM3}.

\subsection{Gust impinging on a NACA0012 aerofoil} \label{sc:naca}

The last example, inspired by~\cite{Sanjay2020}, considers the simulation of a gust impinging on a NACA0012 aerofoil immersed in an incompressible flow at $Re=1,000$. Following~\cite{golubev2009high}, the gust is introduced via a localised source term. The source term in~\eqref{eq:NS} is given by 
\begin{equation} \label{eq:gustSource}
\bm{s}(\bx,t) = 
\begin{cases} 
	\begin{Bmatrix}
		\beta K g(x_1)  \lambda(x_2) \cos\left( \omega t  - \alpha x_1^g \right) \\
		K g'(x_1) \lambda(x_2) \sin\left( \omega t  - \alpha x_1^g \right)
	\end{Bmatrix} &  \text{ if } t \in [50,51] \\
	\bm{0}                                               &  \text{ otherwise }
\end{cases}
\end{equation}
where $(x_1^g,0)$ denotes the centre of the rectangle of dimension $a \times b$ where the gust is generated, the wave number is given by $\alpha = \omega/ v_\infty$ and $v_\infty$ the magnitude of the free-stream velocity. The constant $K$ is defined as
\begin{equation*}
	K = \frac{\left( \alpha^2 - \hat{a}^2 \right) v_\infty^2}{\hat{a}^2  \sin\left(\frac{\displaystyle \alpha \pi}{\displaystyle \hat{a}}\right) }
\end{equation*}
where $\hat{a}$ defines the region where the gust is generated, namely $\hat{a} = 2\pi/a$. Finally, the functions 
\begin{equation}
	\lambda(x_2) = \frac{1}{2} \Big( \tanh \big( 2\pi(x_2 + b/2) \big) - \tanh \big( 2\pi(x_2 - b/2) \big)  \Big)
\end{equation}
and
\begin{equation}
	g(x_1) = 
	\begin{cases} 
		\frac{1}{2}\big( 1 + \cos(\hat{a} (x_1 - x_1^g)) \big)   &  \text{ if } \quad |x_1-x_1^g|\leq \frac{a}{2} \\
		0                                               &  \text{ otherwise } 
	\end{cases}
\end{equation}
are used to guarantee a smooth transition of the flow field in the boundary of the gust region. In the current example, the parameters that define the gust are taken as $a=1$, $b=4$, $x_1^g = 1.52$ and $\omega = 4\pi$.

\begin{figure}[!tb]
	\centering
	\includegraphics[width=0.6\textwidth]{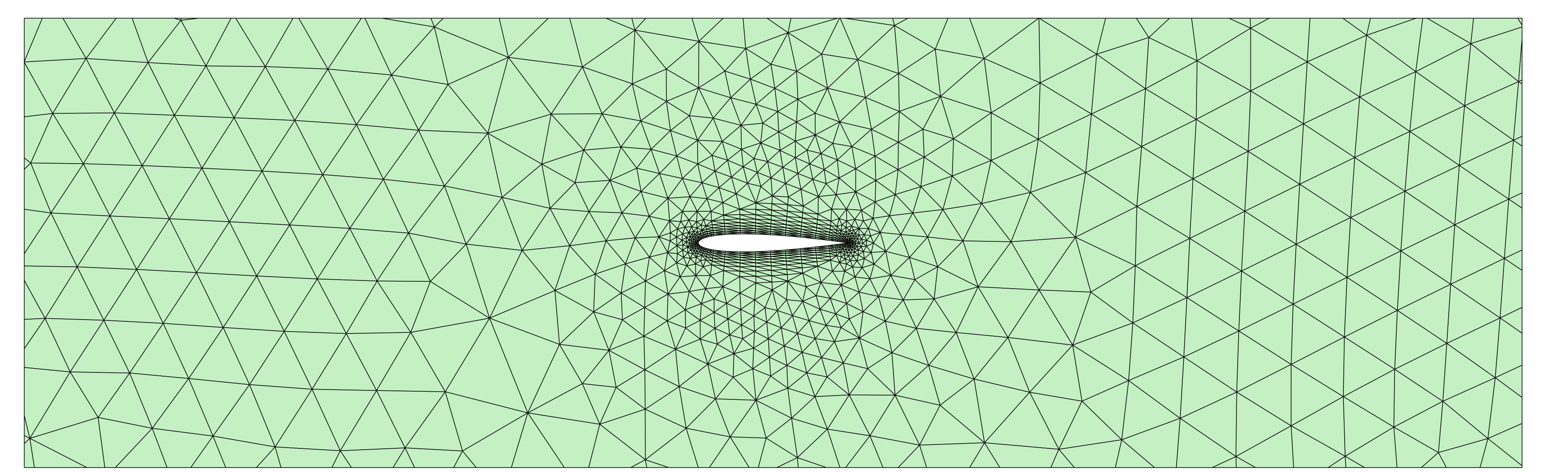}
	\caption{Gust impinging on a NACA0012 aerofoil: detail of the unstructured triangular mesh near the aerofoil.}
	\label{fig:naca_Mesh}
\end{figure}

An unstructured mesh of 2,784 triangles is employed for this example. Curved elements are generated near the aerofoil using the elastic analogy presented in~\cite{xie2013generation}. The size in the normal direction of the first element around the aerofoil is 0.01 and the growing factor in the normal direction is 1.2. Two point sources are introduced to prescribe a mesh size of 0.1 near the leading and trailing edges of the aerofoil, another point source is placed at the centre of the aerofoil to prescribe a size of 0.1 in the vicinity of the aerofoil, whereas a line source with size 0.4 is placed in the path of the gust. A detailed view of the mesh near the cylinders is shown in Figure~\ref{fig:naca_Mesh}.

Given the more complex flow dynamics of this problem, a time step $\Delta t = 0.1$ and the solution is advanced using the ESDIRK46 method until a final time $T = 64$. As commonly done when simulating gust around aerodynamic obstacles~\cite{golubev2009high,Sanjay2020} the initial condition is taken as the steady state solution of the flow around the aerofoil, in this case for $Re=1,000$. The gust is then introduced via the source term and advanced until the final time, selected so that the gust effect in the aerodynamic forces on the aerofoil disappears. 

\begin{figure}[!tb]
	\centering
	\subfigure[$t=50$]{\includegraphics[width=0.48\textwidth]{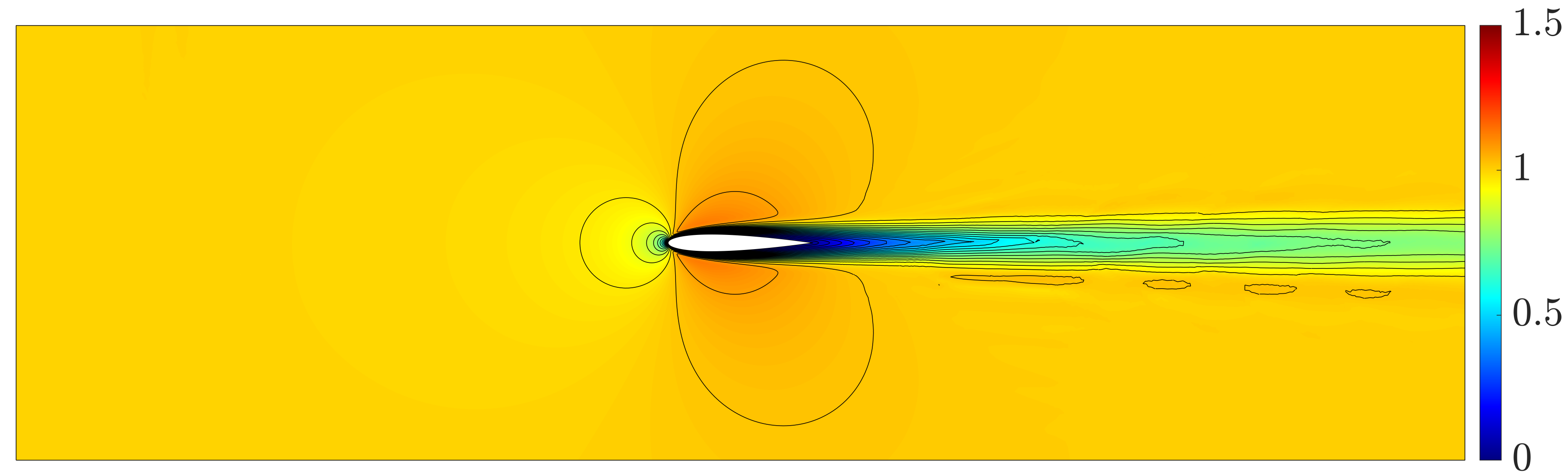}}
	\subfigure[$t=52$]{\includegraphics[width=0.48\textwidth]{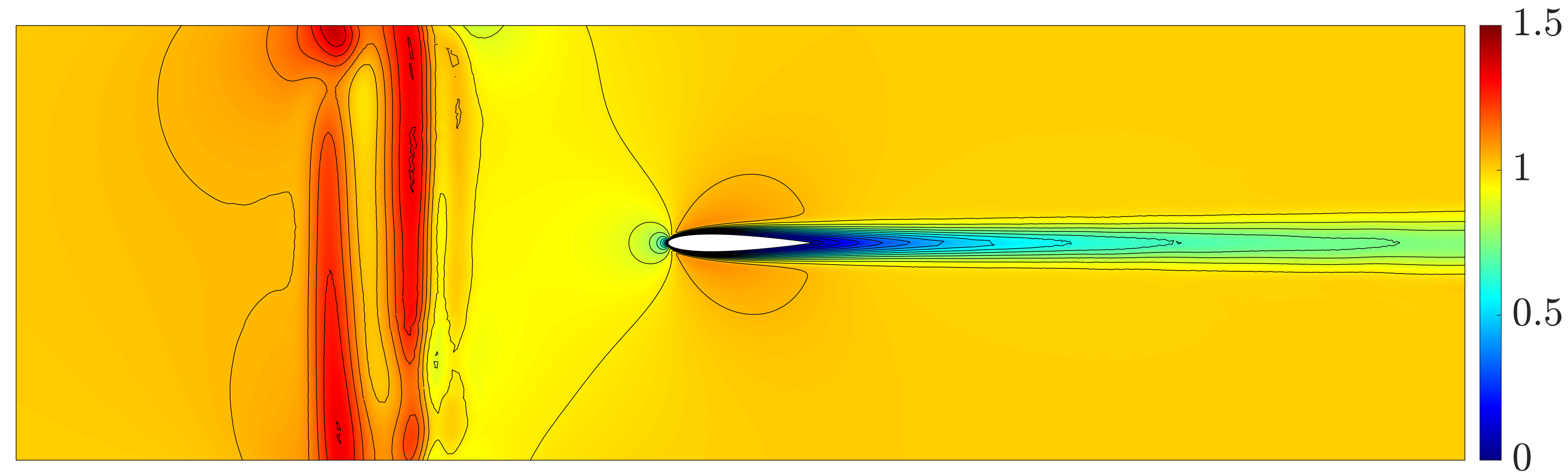}}
	\subfigure[$t=54$]{\includegraphics[width=0.48\textwidth]{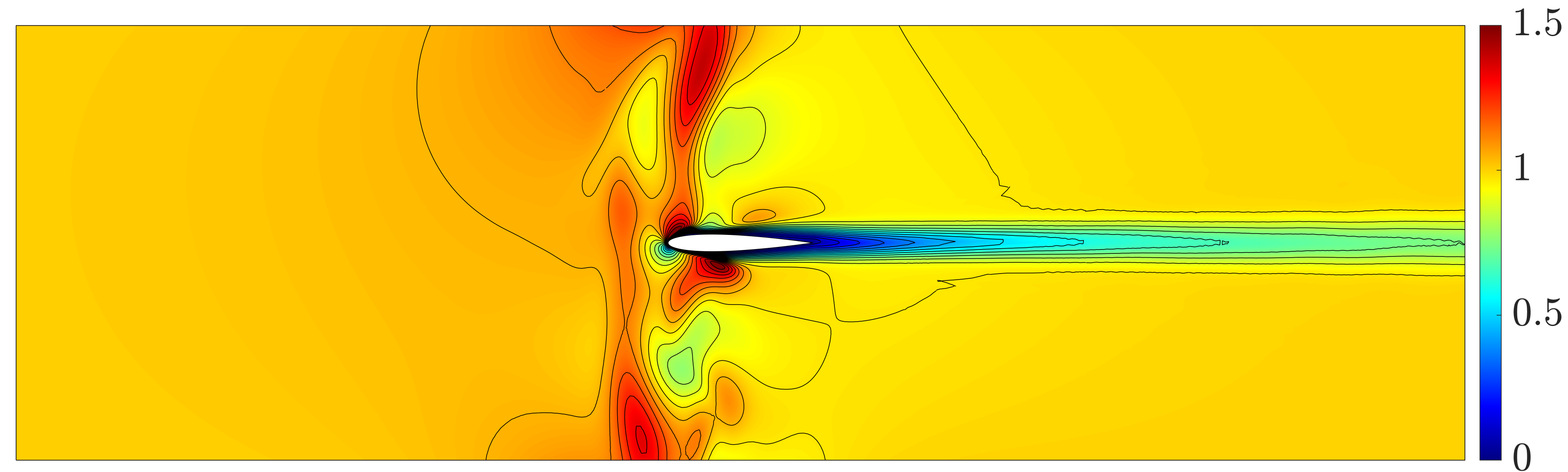}}
	\subfigure[$t=56$]{\includegraphics[width=0.48\textwidth]{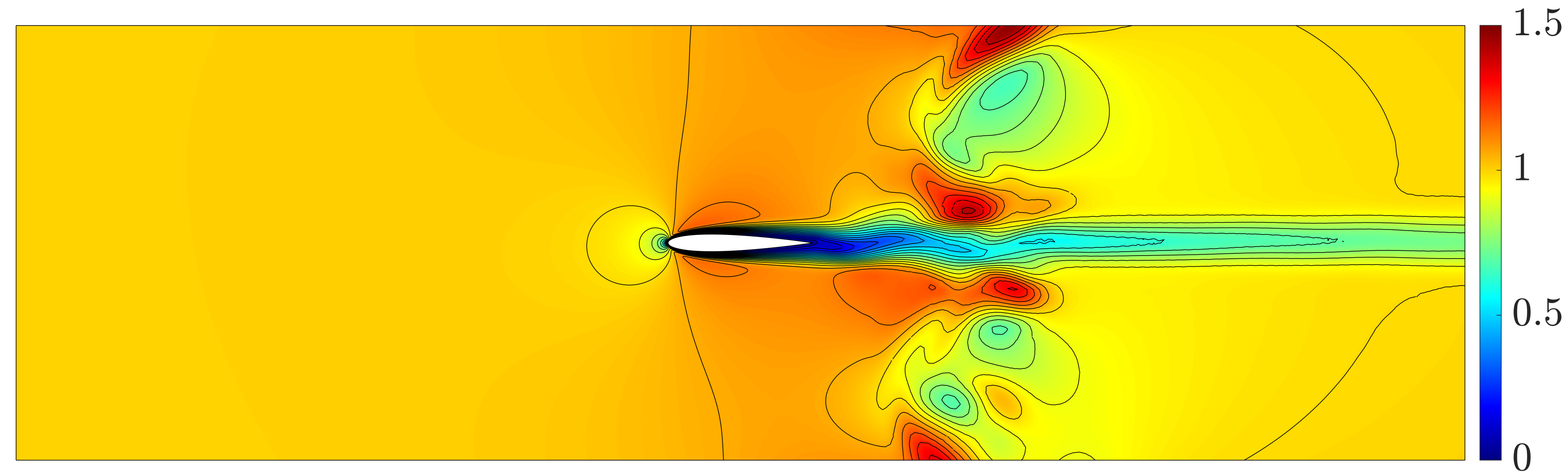}}
	\subfigure[$t=58$]{\includegraphics[width=0.48\textwidth]{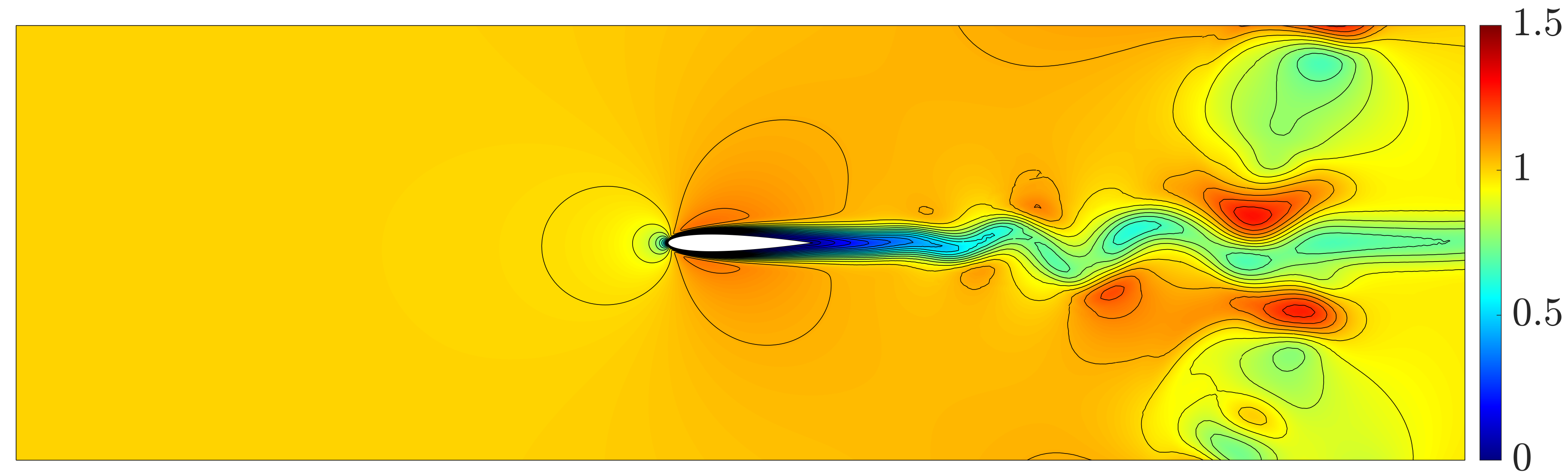}}
	\subfigure[$t=60$]{\includegraphics[width=0.48\textwidth]{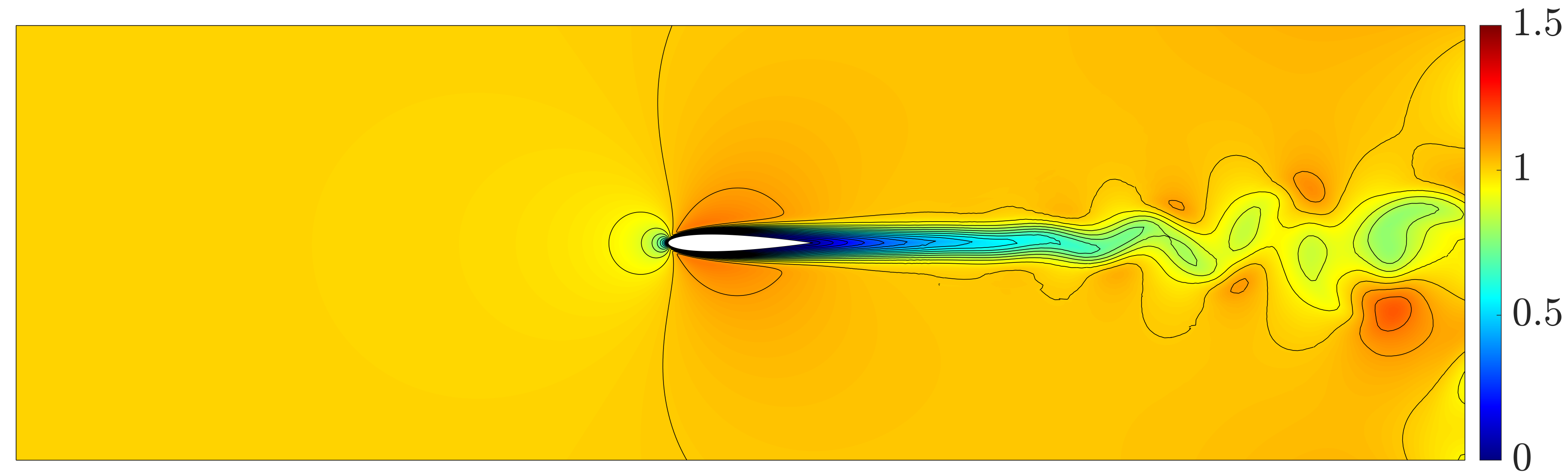}}
	\subfigure[$t=62$]{\includegraphics[width=0.48\textwidth]{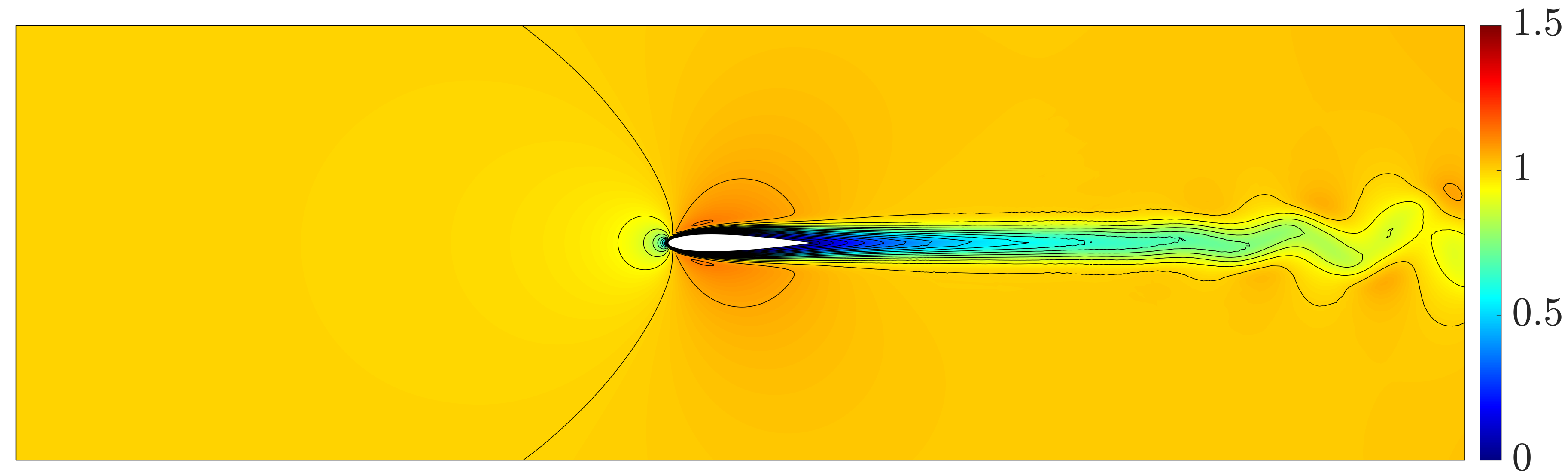}}
	\subfigure[$t=64$]{\includegraphics[width=0.48\textwidth]{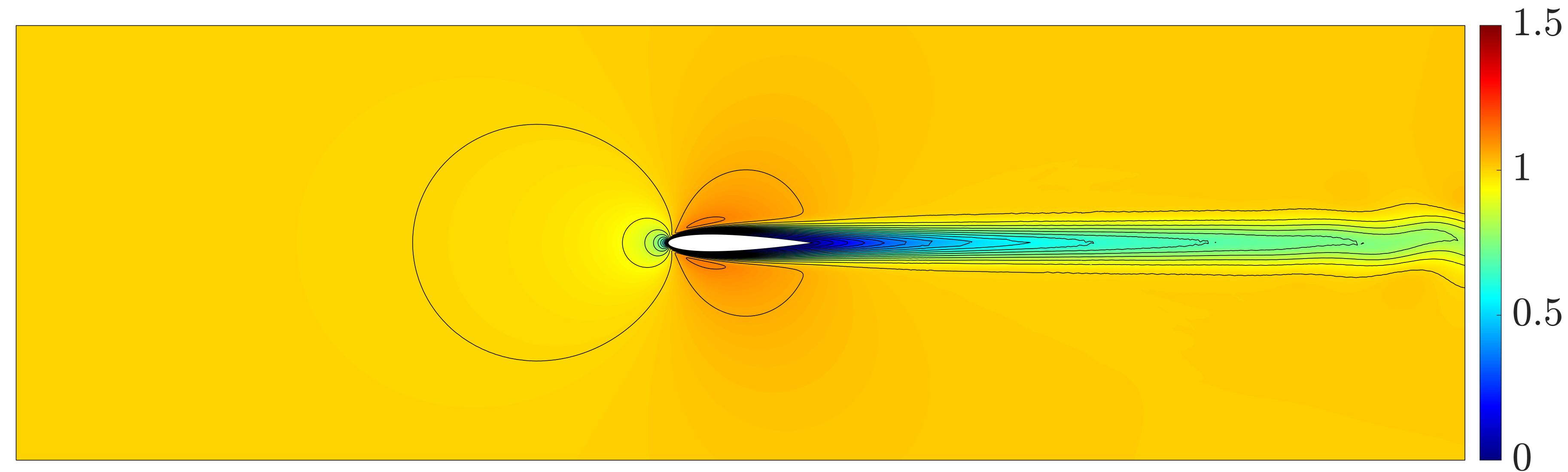}}
	\caption{Gust impinging on a NACA0012 aerofoil: Magnitude of the velocity fields at different instants with a uniform degree of approximation $k=6$.}
	\label{fig:naca_RefVelo}
\end{figure}

As in the previous example, a reference solution is computed by using a uniform degree of approximation $k=6$. The degree of approximation $k=6$ is selected after performing a convergence study on the fixed mesh of Figure~\ref{fig:naca_Mesh}. The magnitude of the velocity at some selected instants is displayed in Figure~\ref{fig:naca_RefVelo}, showing the initial steady state solution, the perturbation of the velocity arriving and impinging on the aerofoil, the complex transient effects induced by the gust and the recovery of the steady state solution after the gust effects disappear.

The need for adaptivity in this example is even more obvious than in the previous example because the perturbation of the velocity is very localised and using a high-order approximation in the whole domain is clearly unnecessary. Next, the standard adaptive process and the adaptivity enhanced with the proposed conservative projection are considered. To remove the effect of the gust generation, when the source term that generates the gust is active, i.e. for $t\leq 10$, a constant degree of approximation $k=6$ is used in both cases. After that time the corresponding adaptive calculation is activated. This ensures that the differences in the adaptive process are not caused by a different representation of the gust. In this example, a desired error of $\varepsilon = 10^{-3}$ is utilised during the adaptive process.

\begin{figure}[!tb]
	\centering
	\subfigure[Lift]{\includegraphics[width=0.48\textwidth]{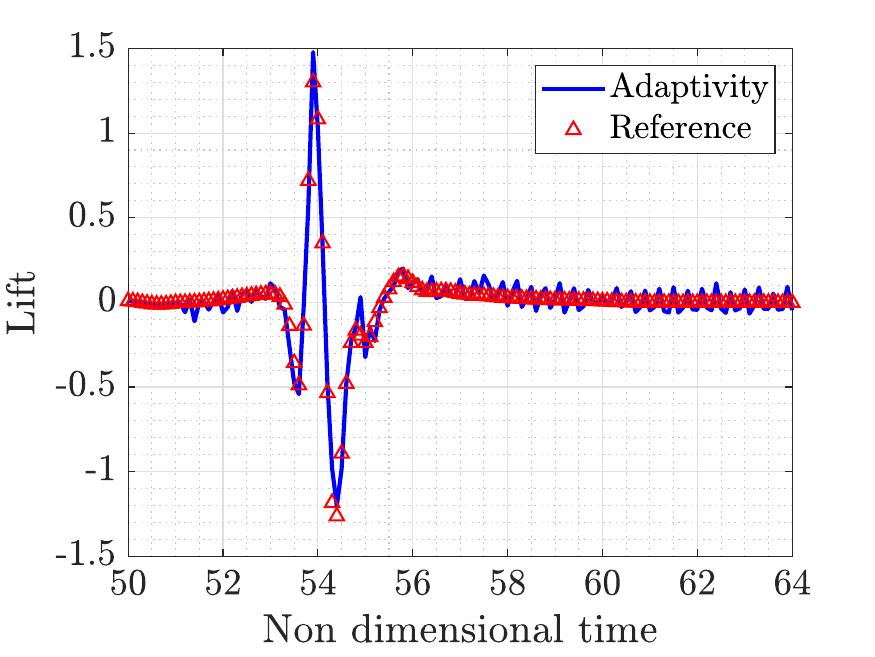}}
	\subfigure[Drag]{\includegraphics[width=0.48\textwidth]{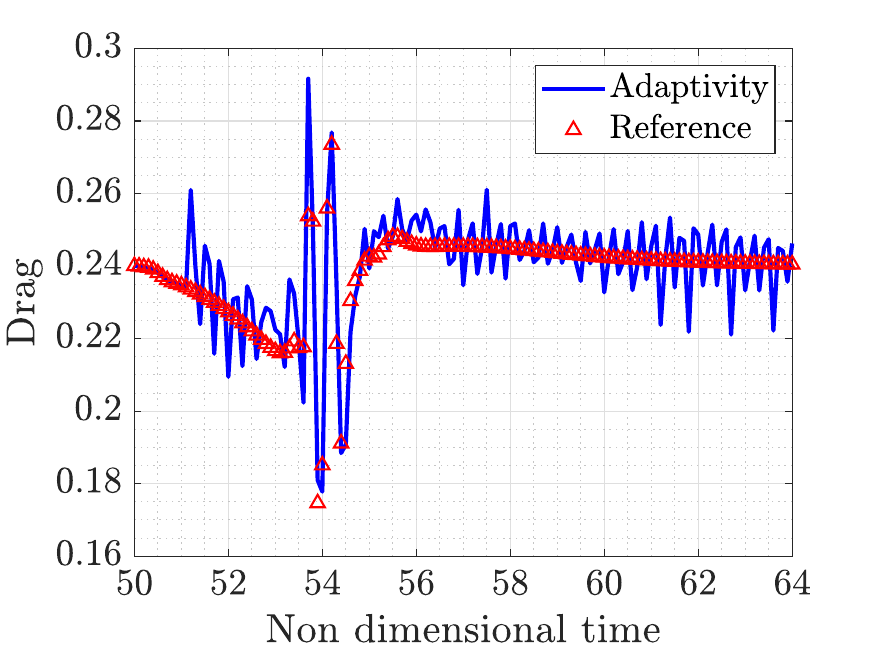}}
	\caption{Gust impinging on a NACA0012 aerofoil: lift and drag using degree adaptivity compared to the reference solution.}
	\label{fig:Naca_NoCorrect}
\end{figure}

Figure~\ref{fig:Naca_NoCorrect} shows the lift and drag on the aerofoil using a standard degree adaptivity and the results are compared to the reference solution. As in the previous example the results show non-physical oscillations. 
The oscillations are more pronounced on the drag but can also be observed on the lift in this example due to the lack of symmetry introduced by the gust. During the transient simulation, a maximum error of  $2.3 \times 10^{-1}$ and $3.8 \times 10^{-2}$ is observed in the lift and drag respectively, clearly not providing the required accuracy for this simulation. It is worth noting that from $t=50$ to $t=51$ a constant degree of approximation, $k=6$, is used and as soon as the adaptivity is activated, a strong overshoot in the drag is observed.

\begin{figure}[!tb]
	\centering
	\subfigure[Lift]{\includegraphics[width=0.48\textwidth]{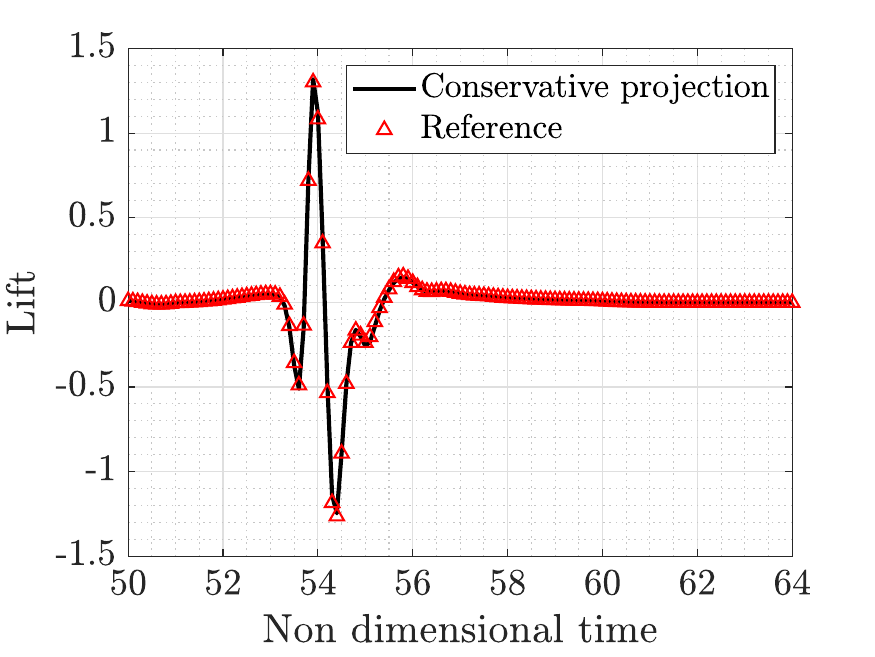}}
	\subfigure[Drag]{\includegraphics[width=0.48\textwidth]{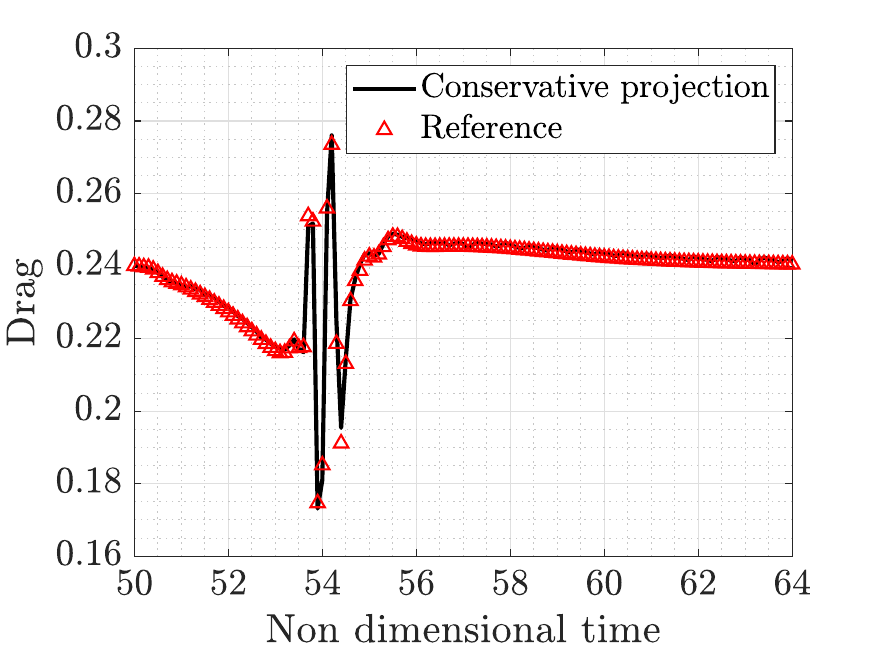}}
	\caption{Gust impinging on a NACA0012 aerofoil: lift and drag using degree adaptivity and the proposed correction compared to the reference solution.}
	\label{fig:Naca_Correct}
\end{figure}

When the proposed correction is introduced, an excellent agreement is again observed between the computed lift and drag and the reference solution, as shown in Figure~\ref{fig:Naca_Correct}.
For this example, the maximum error in the lift and drag during the whole transient process is $5.4 \times 10^{-2}$ and $6.2 \times 10^{-3}$, respectively, showing the extra accuracy provided by the conservative projection of the solution during the adaptive process.

To further quantify the extra accuracy provided by the proposed projection the $\eltwo([51,64])$ norm of the relative lift and drag error is computed for both adaptive approaches. Without the proposed correction the errors in lift and drag are $6.3 \times 10^{-2}$ and $1.4 \times 10^{-3}$ respectively, whereas when the conservative projection is used the errors in lift and drag are more than 40 times lower, namely $1.5 \times 10^{-3}$ and  $2.9 \times 10^{-5}$.

\begin{figure}[!tb]
	\centering
	\subfigure[Velocity, $t=52$] {\includegraphics[width=0.48\textwidth]{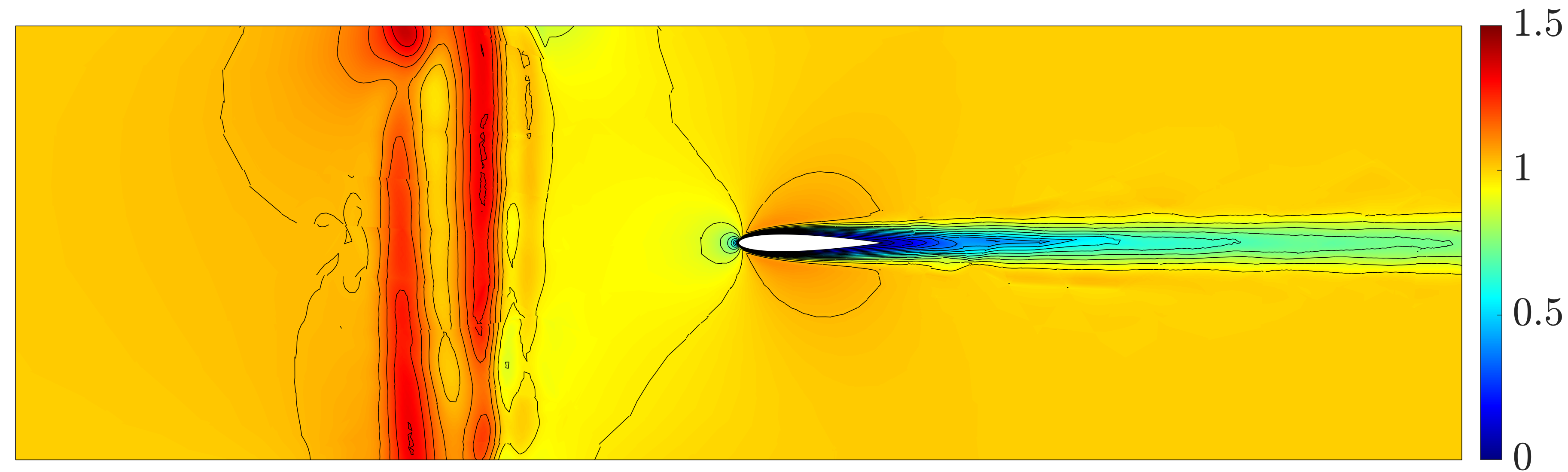}}
	\subfigure[Degree map, $t=52$]{\includegraphics[width=0.48\textwidth]{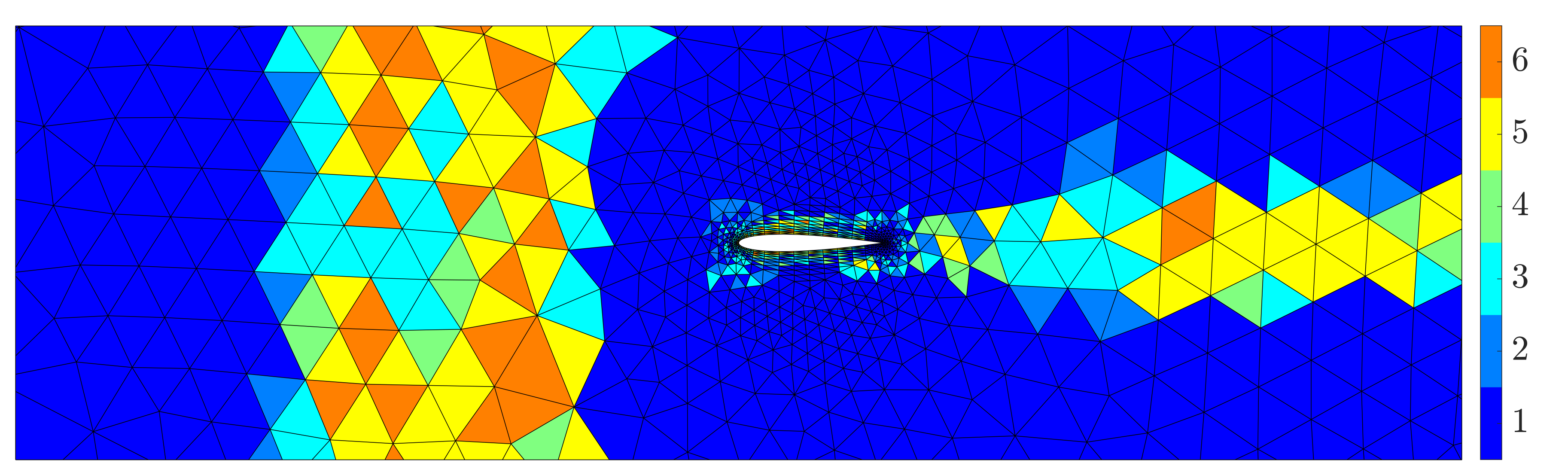}}
	\subfigure[Velocity, $t=54$] {\includegraphics[width=0.48\textwidth]{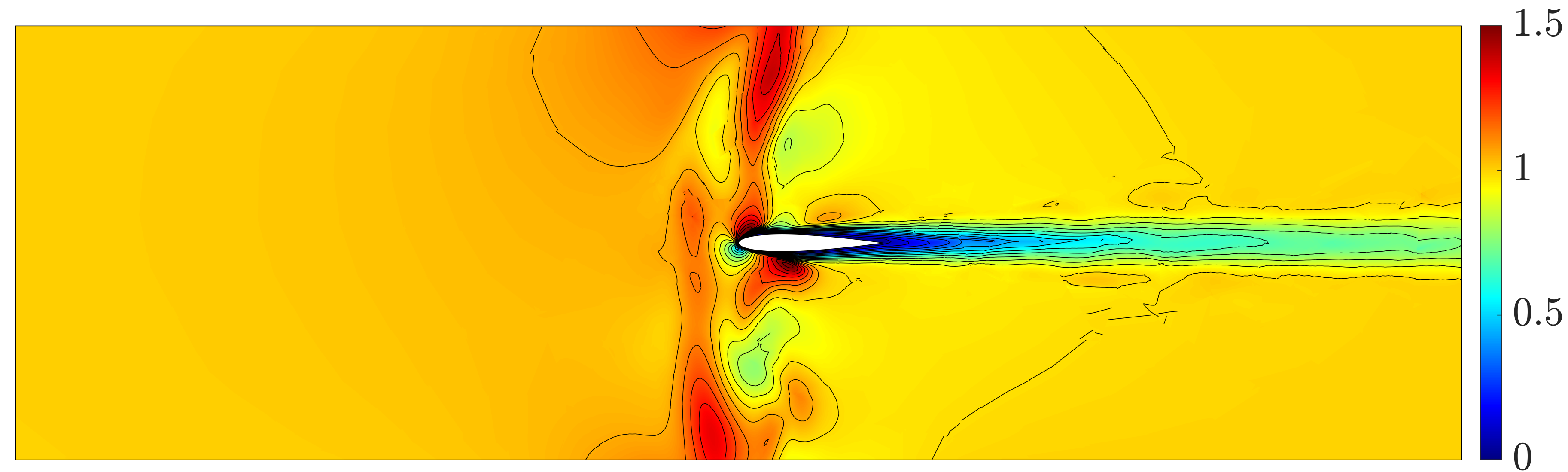}}
	\subfigure[Degree map, $t=54$]{\includegraphics[width=0.48\textwidth]{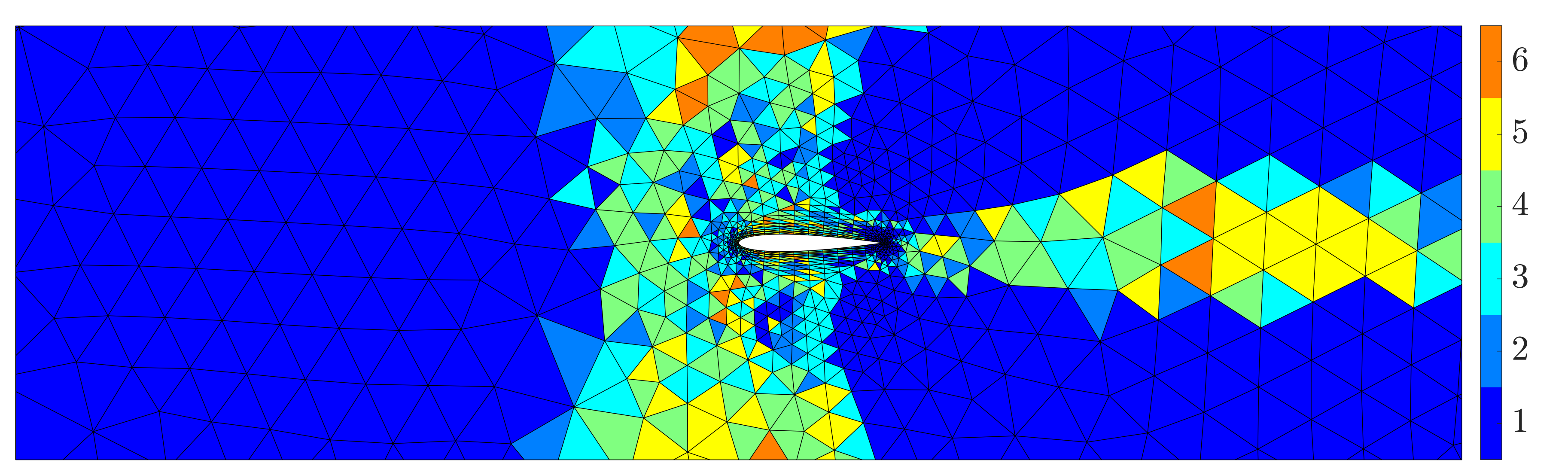}}		
	\subfigure[Velocity, $t=56$] {\includegraphics[width=0.48\textwidth]{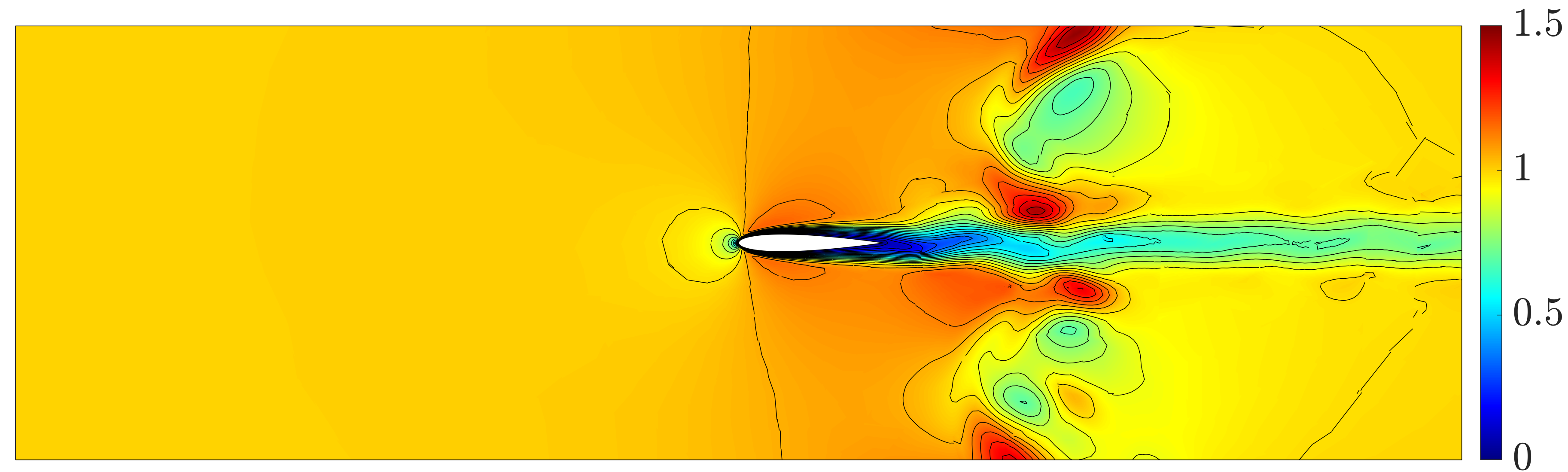}}
	\subfigure[Degree map, $t=56$]{\includegraphics[width=0.48\textwidth]{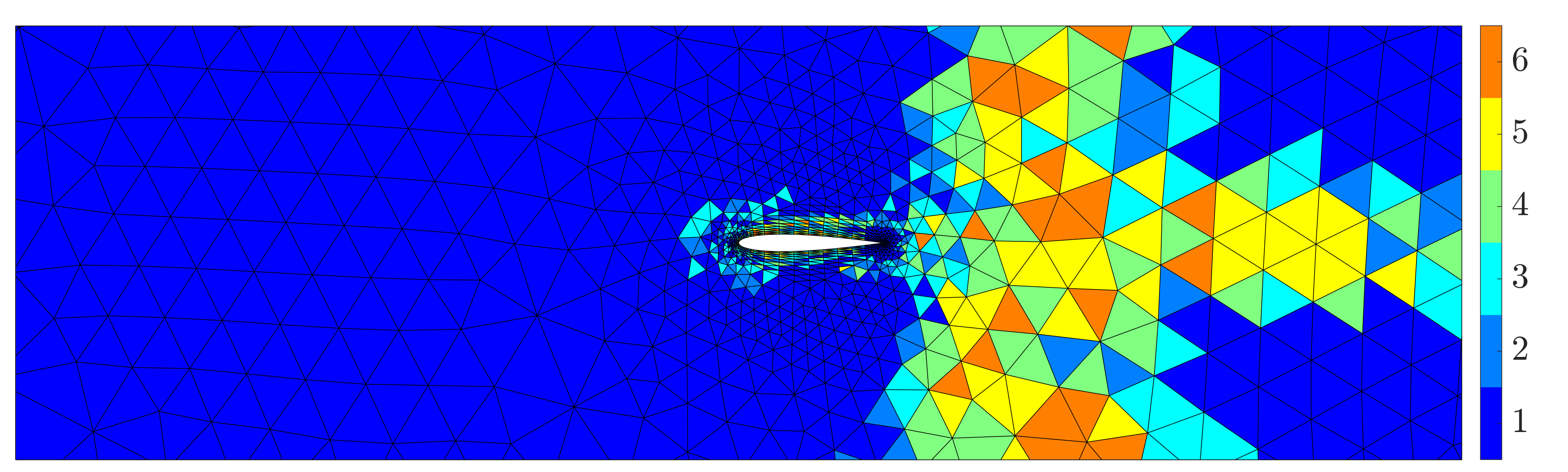}}	
	\subfigure[Velocity, $t=58$] {\includegraphics[width=0.48\textwidth]{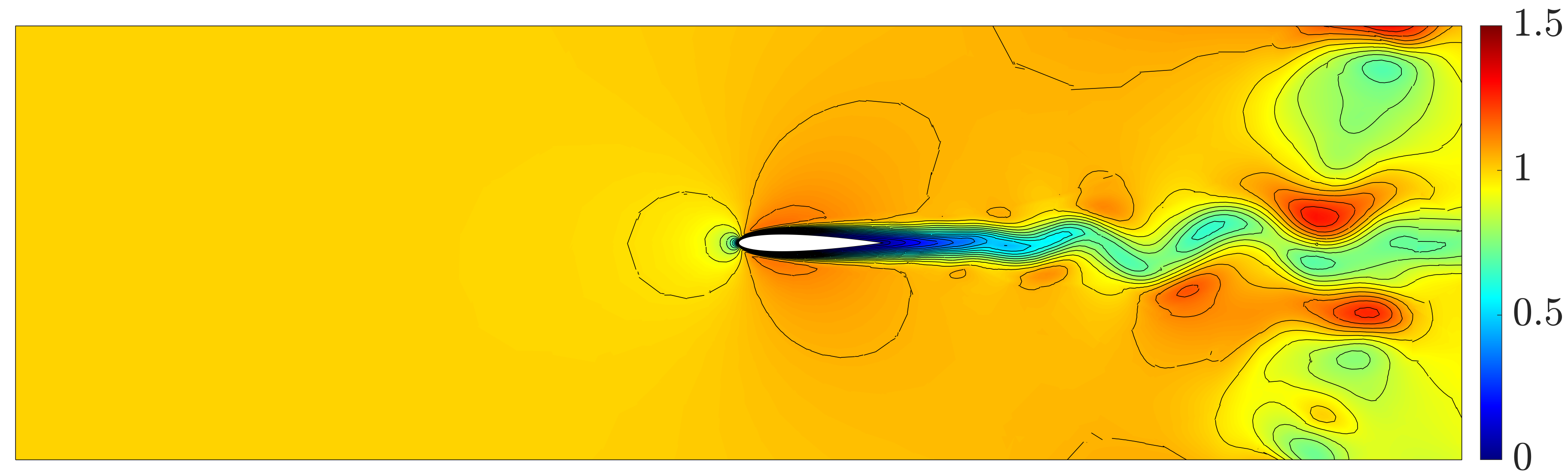}}
	\subfigure[Degree map, $t=58$]{\includegraphics[width=0.48\textwidth]{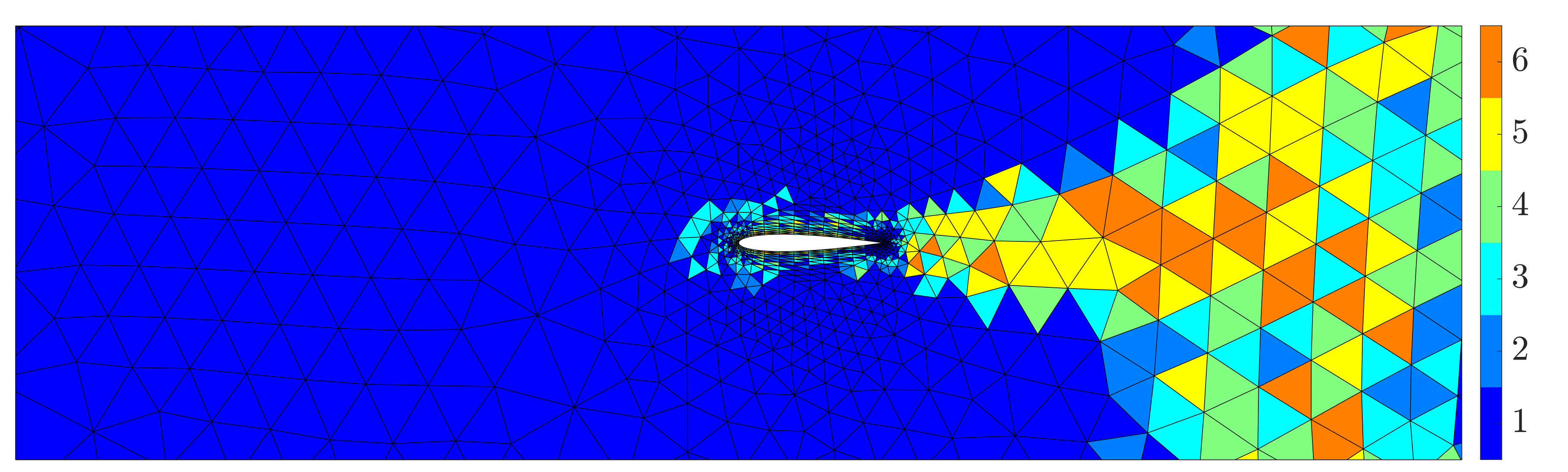}}	
	\subfigure[Velocity, $t=60$] {\includegraphics[width=0.48\textwidth]{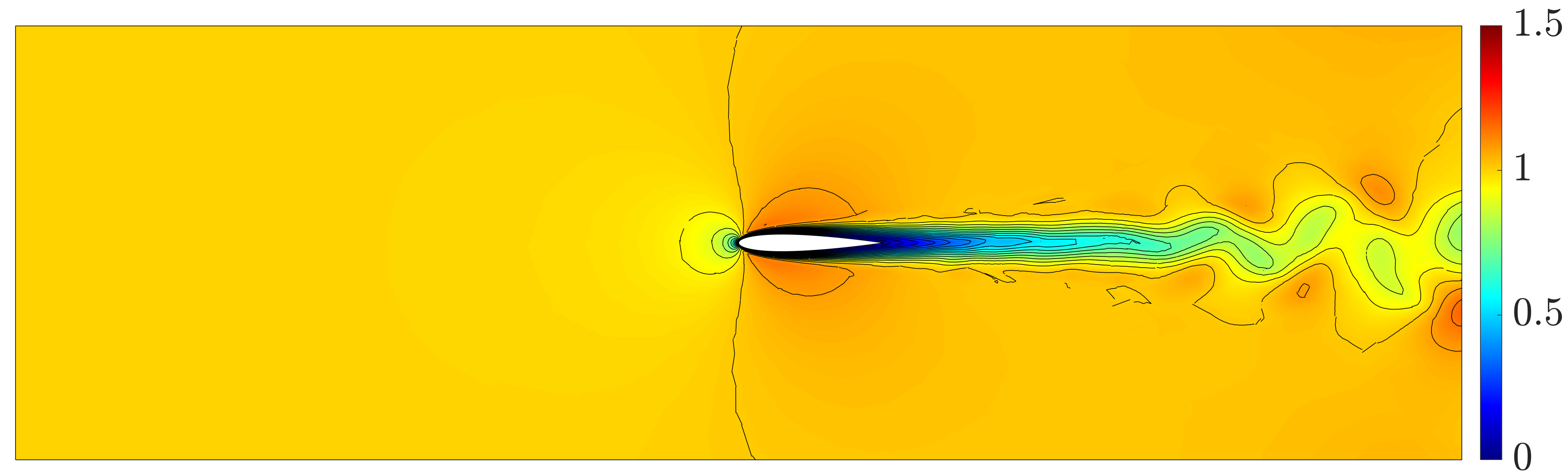}}
	\subfigure[Degree map, $t=60$]{\includegraphics[width=0.48\textwidth]{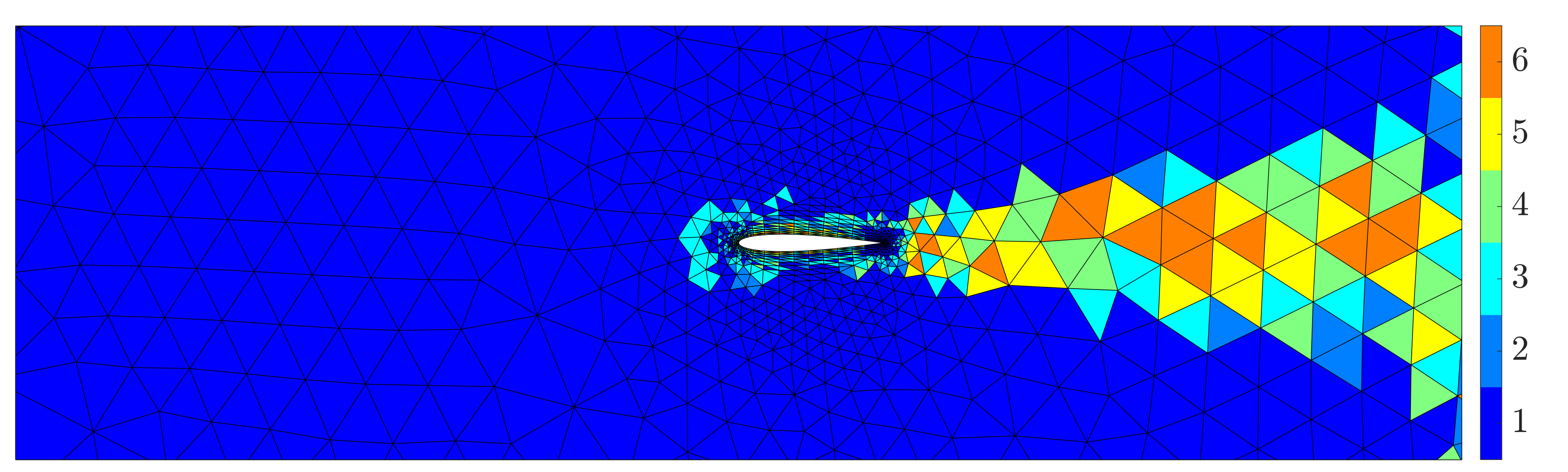}}	
	\caption{Gust impinging on a NACA0012 aerofoil: Magnitude of the velocity fields (left) and map of the degree of approximation (right) at different instants with the proposed degree adaptive approach.}
	\label{fig:naca_AdaptivityVeloPmap}
\end{figure}

To illustrate the ability of the degree adaptive process to accurately capture the complex flow features of this problem, lowering the degree on the elements where accuracy is no longer required, Figure~\ref{fig:naca_AdaptivityVeloPmap} shows the magnitude of the velocity and the degree map at some selected instants. 
Comparing the results with the reference solution of Figure~\ref{fig:naca_RefVelo}, it can be observed that the adaptive process captures all the flow features. The degree map clearly reflects the regions where the complexity of the solution requires a higher degree of approximation to provide the desired accuracy. 

\begin{figure}[!tb]
	\centering
	\includegraphics[width=0.6\textwidth]{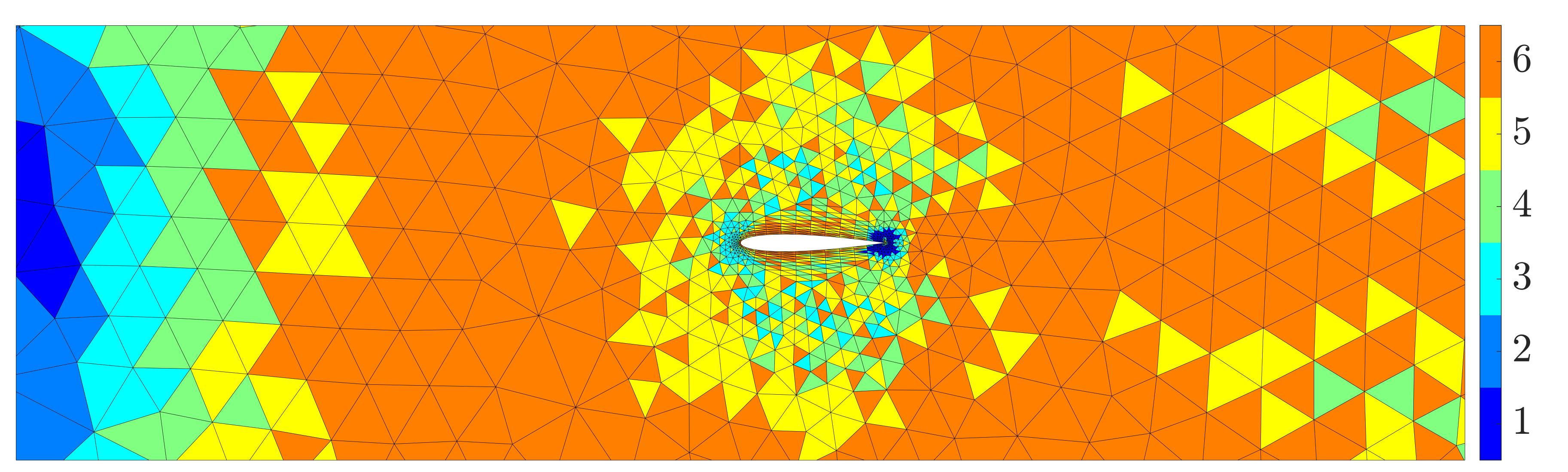}
	\caption{Gust impinging on a NACA0012 aerofoil: Map of the degree of approximation at $t=64$ with an adaptive process not allowing the degree to be lowered.}
	\label{fig:naca_pMapNotLowering}
\end{figure}

\begin{figure}[!tb]
	\centering
	\includegraphics[width=0.6\textwidth]{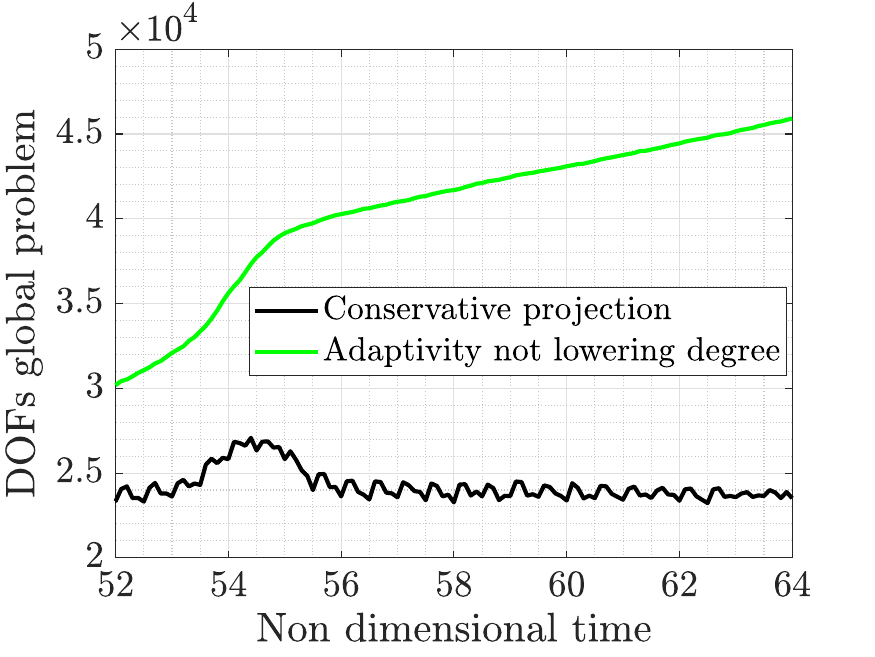}
	\caption{Gust impinging on a NACA0012 aerofoil: Number of degrees of freedom of the global problem for two different adaptive approaches.}
	\label{fig:naca_DOFs}
\end{figure}

In this example, the ability to lower the degree of approximation is critical to gain the benefits of a degree adaptive process, without compromising the accuracy. As the gust introduces a localised perturbation of the velocity, without lowering the degree the final degree map shows that a high order polynomial approximation is used in many areas where the flow does not show any feature. The degree map for such an approach is displayed in Figure~\ref{fig:naca_pMapNotLowering}.
To quantify the benefit of the proposed conservative projection, Figure~\ref{fig:naca_DOFs} show the number of degrees of freedom of the global problem as a function of the non-dimensional time for the proposed approach and an adaptive process where the degree is not allowed to be decreased during the time marching process.
With the proposed projection the number of degrees of freedom at $t=64$ is 23,518 whereas for the approach not lowering the degree of approximation the number of degrees of freedom at $t=64$ reaches 45,908. The results with the conservative projection show that the most complex dynamics happen at around $t=54$, which, according to Figure~\ref{fig:naca_AdaptivityVeloPmap}, is precisely when the gust impinges on the aerofoil. At this point the number of degrees of freedom of the global problem reaches a maximum and then decreases because the degree of approximation can be lowered in many elements in the vicinity of the aerofoil where the transient effects are no longer relevant.

In terms of computational cost, the simulation with the proposed conservative projection is more than three times faster than the simulation with a uniform degree of approximation $k=6$. The extra performance compared to the previous example is due to the localised effect of the gust. In this example, the degree adaptive clearly offers a major advantage by introducing high order approximation only where needed.

\section{Concluding remarks} \label{sc:conclusions}

A new conservative projection has been proposed and tested within the context of degree adaptivity for the solution of transient incompressible Navier-Stokes flows. Without this projection, a standard degree adaptive process leads to non-physical oscillations in the aerodynamic quantities of interest when the degree of approximation is lowered during the time marching process. These oscillations are linked to the violation of the incompressibility condition when the degree of approximation is lowered, leading to oscillations in the pressure field. To provide further evidence about the nature of these oscillations, an adaptive process where the degree of approximation is not allowed to be lowered during the time marching has been implemented, leading to correct solutions. However, the extra cost of this approach makes the adaptivity not an efficient choice, especially in problems where localised transient effects travel along the domain.

The proposed conservative projection completely removes the non-physical oscillations in the aerodynamic quantities of interest and enables the degree to be lowered in regions where accuracy is no longer required, leading to a more efficient use of high order approximations, only where needed.

Two examples have been used to illustrate the benefits of the proposed approach and to quantify the extra accuracy and the lower computational requirements compared to a standard degree adaptive approach and to an adaptive strategy where the degree is not allowed to be lowered.

\bibliographystyle{ieeetr}
\bibliography{conservativePAdaptation}

\end{document}